\pdfoutput=1
\PassOptionsToPackage{dvipsnames}{xcolor}
\documentclass[longbibliography,showpacs,floatfix,superscriptaddress,twocolumn,aps,prl]{revtex4-2}
\usepackage[figuresright]{rotating}
\usepackage{amssymb}
\usepackage{bm}
\usepackage{amsmath}
\usepackage{mathtools}
\usepackage{psfrag}
\usepackage{multirow}
\usepackage{tabularx}
\usepackage{textcomp}
\usepackage{units}
\usepackage{lipsum}
\usepackage{wasysym}
\usepackage{graphicx}%

\usepackage{physics}
\usepackage{soul}
\usepackage{titlesec}
\usepackage{times}

\usepackage{enumitem}

\usepackage[caption=false]{subfig}

\usepackage{tikz}
\usepackage[colorlinks=true,citecolor=RubineRed,linkcolor=RubineRed,urlcolor=RubineRed,filecolor=RubineRed]{hyperref}

\makeatletter
\let\frontmatter@footnote@produce\frontmatter@footnote@produce@footnote
\makeatother

\newcommand{\br}{\textbf r}
\newcommand{\bR}{\textbf R}
\newcommand{\bq}{\textbf q}
\newcommand{\bk}{\textbf k}

\newcommand{\ii}{\mathrm i}
\newcommand{\sgn}{\mathrm{sgn}}

\begin{document}

\title{Bulk--Defect Correspondence in Algebraic Classical Spin Liquids}

\newcommand{\TUM}{\affiliation{Technical University of Munich, TUM School of Natural Sciences, Physics Department, 85748 Garching, Germany}}
\newcommand{\MCQST}{\affiliation{Munich Center for Quantum Science and Technology (MCQST), Schellingstr. 4, 80799 M{\"u}nchen, Germany}}

\author{Rafael Flores-Calder\'on}
\email[Electronic address: ]{r.flores-calderon@tum.de}
\TUM \MCQST

\date{\today}

\begin{abstract}
Local constraints in algebraic classical spin liquids give rise to singular momentum-space correlations. This so called pinch-point singularity carries a momentum-space winding number that determines the low-energy structure of the algebraic spin liquid. We show that nonmagnetic vacancies in two dimensions reveal this topological invariant in the real-space winding of the surrounding spin textures. Under explicit conditions, we prove this bulk--defect correspondence for rank-$n$ $\mathrm{U}(1)$ spin liquids and verify it numerically for the Coulomb spin liquids of the kagome and checkerboard lattices and the rank-2 $\mathrm{U}(1)$ spin liquid of the honeycomb-snowflake model. In this way we show that vacancies can be used as general probes for distinguishing the emergent gauge theories of strongly correlated classical disordered magnets.
\end{abstract}

\maketitle

\begin{figure}[t]
\centering
\includegraphics[width=\columnwidth]{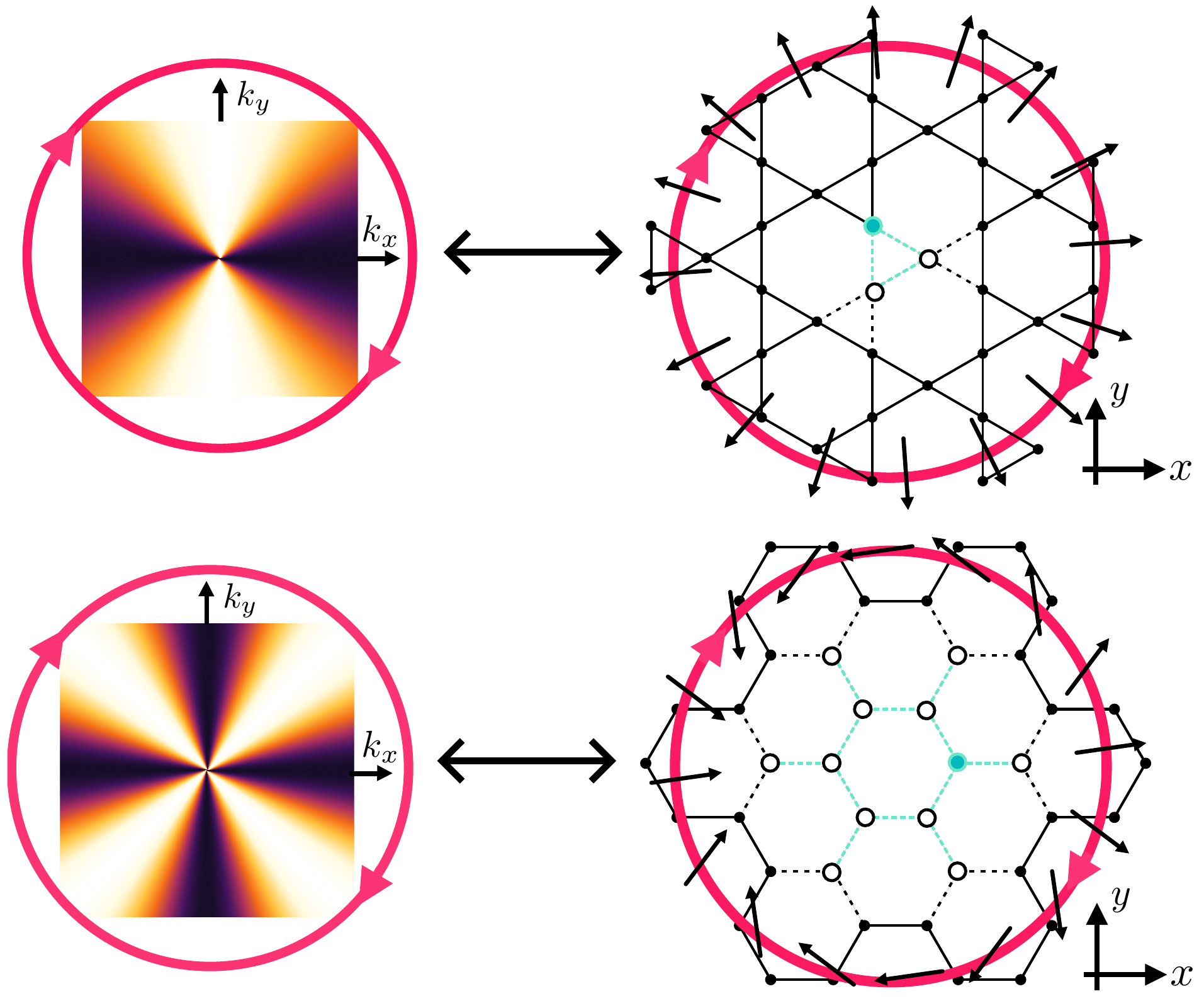}
\caption{Bulk--defect correspondence at rank 1 (top, kagome) and rank 2 (bottom, honeycomb snowflake).  Left: pinch point around the band touching, two-fold at rank 1 and four-fold at rank 2; the winding of the constrainer on the red loop is $W_{\rm bulk}$.  Right: spin texture (arrows) around the orphan spin (filled turquoise circle) left by a vacancy cluster (open circles); the turquoise dashed lines mark the cluster of the orphan's constrainer.  On the red circle the texture winds $W_{\rm defect}=W_{\rm bulk}$ times, once at rank 1 and twice at rank 2.}
\label{fig:schematic}
\end{figure}

\paragraph{Introduction.---}

Spin liquids are strongly interacting magnets without conventional order, instead characterized by emergent gauge fields and fractionalized excitations~\cite{Anderson73,Wen02,Balents10,Castelnovo12,Zhou17,KnolleMoessner19}. In classical spin liquids, local constraints select an extensively degenerate ground-state manifold~\cite{MoessnerChalker98PRL,MoessnerChalker98PRB} and become Gauss laws of an emergent gauge theory upon coarse graining~\cite{Isakov04,Henley05,Henley10}. These constraints leave a characteristic signature in momentum space: pinch points, whose singular angular structure is visible in neutron scattering~\cite{Fennell09}. This description extends beyond conventional Coulomb phases to higher-rank gauge theories, in which symmetric tensor fields replace the familiar vector field~\cite{Xu06,XuFisher07,XuHorava10,Pretko17,Pretko17b,Prem18}. Their generalized Gauss laws can impose additional conservation laws, such as dipole conservation, restricting the motion of charges; fully immobile charges are known as fractons~\cite{Chamon05,Haah11,Vijay15,Vijay16,Pretko17,Halasz17,YouvonOppen19,NandkishoreHermele19,PretkoChenYou20}. Interest in these theories also stems from their connections to fault-tolerant quantum computation~\cite{KubicaYoshida18,Schmitz18} and elasticity~\cite{PretkoRadzihovsky18}. Several classical spin models now realize higher-rank gauge fields~\cite{YanClassification,Benton16,YanRank2,BentonMoessner21}, providing microscopic settings in which to explore their consequences. The constrainer formalism developed in Refs.~\cite{BentonMoessner21,YanClassification,YanTypology} brings these phases into a common microscopic framework, relating their generalized Gauss laws to touchings between flat and dispersive bands

Such band touchings can also carry topological information. In electronic systems, boundaries and defects expose bulk topological invariants~\cite{TeoKane10}; in frustrated hopping models, flat-band touchings likewise reflect real-space topology~\cite{Bergman08}. For classical spin liquids, the topology of the constrainer provides a route to new Coulomb phases~\cite{BentonMoessner21}. This raises a complementary question: can a local defect reveal the topological index of a pinch point?

Vacancies offer a direct and experimentally relevant way to address this question. For continuous spins the relevant dilution cluster leaves a single spin in a constraint cluster, the surviving so called orphan spin cannot satisfy the local constraint and acquires an effective fractional moment~\cite{MoessnerBerlinsky99,sen,*SenDamleMoessner11,SenDamleMoessner12,LaForge13,SenMoessner15,Rehn15,Patil20}. This mechanism accounts for anomalous magnetic responses in diluted frustrated magnets~\cite{SchifferDaruka97,Limot02}. The orphan acts as a gauge charge surrounded by a spin texture shaped by the correlations of the host liquid~\cite{SenDamleMoessner12}. Higher-rank constraints related to fractonic excitations can produce irrational orphan moments and nondecaying textures~\cite{MoessnerSondhi10,FloresCalderonBentonMoessner24}, suggesting that defects encode more than the strength of the local response.

Here we establish a bulk--defect correspondence that makes this connection precise. For an isolated algebraic band touching with a real two-component charged sector, we define an integer index $W_{\rm bulk}$ as the vorticity of the constrainer zero. Under explicit conditions, this index equals the real-space winding of the vacancy-induced texture (Fig.~\ref{fig:schematic}). This relation unifies rank-1, rank-2, and higher-rank descriptions with the orphan setting the charge and hence the response amplitude, while the undiluted liquid determines the spatial structure.  Monte Carlo simulations of the kagome and checkerboard Coulomb liquids and the rank-2 honeycomb-snowflake model quantitatively verify the correspondence. A local defect thereby serves as an indicator of the momentum-space pinch-point topology.

\paragraph{Constrainer formalism and the rank-$n$ bulk invariant.---}
Consider a two-dimensional lattice of spins $\vec{S}^a_\textbf{R}$ with $\mathcal{N}$ spin components, where $i=(\textbf{R},a)$ labels sites organized into unit cells $\textbf{R}$ containing $a=1,\dots,N$ sublattices, subject to $\vec{S}_i^2=1$.  Classical spin liquids (CSL) have a highly constrained ground-state manifold which may arise from a Heisenberg Hamiltonian of \textit{constrainer} form~\cite{YanClassification}
\begin{align}
\label{eq:constrainer-H-prl}
\mathcal{H} = \frac{J}{2}\sum_{\textbf{R}} \left[\sum_i C^i(\textbf{R})\vec{S}_i \right]^2,
\end{align}
with a constrainer function $C^i(\textbf{R})$ finite for $i$ in a cluster surrounding $\textbf{R}$.  We consider one constrainer per unit cell, which contains the essential physics; the general case of $N_c$ constrainers per cell, which includes kagome, is treated in the Supplemental Material (SM).  The ground state satisfies $\sum_i C^i(\textbf{R})\vec{S}_i =0$ for every $\textbf R$.
In momentum space $T^{a}(\textbf{q}) = \sum_i e^{-\ii\textbf{r}_i\cdot \textbf{q}}C^a(0,\textbf{r}_i)$, and we set $J=1$ from now on.  The constrainer vector $\textbf T(\bq)=(T^1,\dots,T^N)$ gives $N-1$ flat bands and one dispersive band $\omega(\bq)=|\textbf T(\bq)|^2$, and the ground-state constraint becomes
\begin{align}
\label{eq:ground-constraint-prl}
\sum_a T^a(\bq)^* \vec{S}^a(\bq)=0 , \forall \bq .
\end{align}
A pinch point occurs where the dispersive band touches the flat bands, at a momentum $\bq_0$ with $\textbf T(\bq_0)=0$; algebraic spin liquids have at least one such touching.  We assume throughout that the lattice is inversion symmetric. The reality of the spins together with inversion, see SM, leads to $
T^a(\bq)^{*}=T^{\bar a}(\bq)$, with $\bar a$ the sublattice related to $a$ by inversion $P$.
This is the statement that the antiunitary $\Theta=P\mathcal K$, with $\mathcal K$ complex conjugation, leaves every $\bq_0$ fixed; $\Theta^2=P^2\mathcal K^2=\mathbb I$.
For pinch points with $2\bq_0$ a reciprocal-lattice vector, reality of the spins alone already makes the constrainer real; inversion is needed to pair $+\bq_0$ with $-\bq_0$ when the two are distinct, as at the Brillouin-zone corners $\pm\bm K$ of the honeycomb lattice.  Near the touching the constrainer vanishes as $\textbf T(\bq_0+\bk)=O(k^n)$, which defines the rank $n$.  It can then be written in terms of a vector $d_A(\bk)$ with $A=1,\dots,N$, defined by $\sum_a T^a(\bq_0+\bk)^* \vec{S}^a(\bq_0+\bk)=d_A(\bk)\vec{\Phi}_A(\bk)+\cdots=0$, with Einstein summation over $A$, where the \emph{soft modes}
\begin{equation}
\label{eq:soft-modes-prl}
 \vec\Phi_A(\bk)=\sum_a\mathcal U_{Aa}\,\vec S^a(\bq_0+\bk)
\end{equation}
project the spins onto an orthonormal basis $\mathcal U$ of the $N$ spin modes at the touching, all of which cost no energy at $\bq_0$, chosen such that $T^a(\bq_0+\bk)^*=d_A(\bk)\,\mathcal U_{Aa}+O(k^{n+1})$.  The soft modes are the long-wavelength spin fluctuations, and not all $N$ of them enter the constraint.  We focus on a single pinch-point with a dispersive band and a flat band forming a two-dimensional space, the charged sector. Under this assumption the vectors $\textbf d(\bk)=(d_1(\bk),d_2(\bk),0,\dots,0)$ span a $\Theta$ invariant real two-plane, and we order $\mathcal U$ so that it is spanned by $A=1,2$.  Here $\vec\Phi_1$ and $\vec\Phi_2$ are the charged soft modes, while $\vec\Phi_A$ with $3\le A\le N$ are neutral modes that the constraint does not see and that play no role below.  From here on $A,B=1,2$, and $\textbf d(\bk)$ denotes the two-component vector of non-zero entries. We can now define the bulk invariant as the integer winding number
\begin{equation}
 W_{\rm bulk}=\frac{1}{2\pi}\oint_{\mathcal C}\dd k_\mu\;
 \epsilon_{AB}\,\hat d_A \frac{\partial}{\partial k_\mu}\hat d_B,\label{eq:general-framework-prl}
\end{equation}
 where we defined the real unit vector $\hat{\textbf d}(\bk)=\ii^{-n}\textbf d(\bk)/|\textbf d(\bk)|$, summation over $A,B$ is implied and $\mathcal C$ is any closed contour encircling $\bq_0$.
The invariant is independent of  the $\mathcal C$ parametrization, and it is unchanged by any deformation of the contour encircling $\bq_0$.  It is the vorticity of the pinch-point at $\bq_0$~\cite{BentonMoessner21,YanClassification}, a property of the low-energy theory.  Its magnitude, $|W_{\rm bulk}|$ is basis independent, whereas its sign depends on the orientation of the basis $(\Phi_1,\Phi_2)$. Changing the basis may flip the sign of $W_{\rm bulk}$ together  with the defect topological invariant defined below together, and we fix the convention in the SM.
In real space the local constraint of a rank-$n$ spin liquid leads in the continuum to a generalized Gauss law,
\begin{equation}
\partial_{\mu_1}\cdots\partial_{\mu_n}B_{\mu_1\cdots\mu_n}=0, \label{eq:rankn-gauss-prl}
\end{equation}
for an emergent gauge field $B$, a symmetric traceless rank-$n$ tensor carrying the spin components as spectator labels.  In two dimensions such a tensor has exactly two independent components at every rank, which we take as $B_1=B_{xx\cdots x}$ and $B_2=B_{xx\cdots xy}$.
The two charged soft modes are the two independent components of the gauge field, $\vec B_A=L_{AB}\vec\Phi_B$, with a constant invertible $2\times2$ matrix $L$ fixed by the constrainer.

\paragraph{Vacancy spin textures and gauge charges.---}
Let us now consider the spin system in the presence of nonmagnetic vacancies.  For continuous spins, such as Heisenberg spins ($\mathcal{N}=3$), removing a single spin does not modify the low-energy physics: the surrounding spins can reorient and still satisfy the local constraints of the ground state,similarly for more spins removed. If, however, all the spins of one constraint cluster are removed except one, that constraint can no longer be satisfied, we have
\begin{equation}
 \vec{C}_0\equiv\sum_i C^i(\textbf{R}_0)\vec{S}_i =m_{\rm O}\,\vec n\neq 0,
\label{eq:orphan-is-charge-prl}
\end{equation}
where $\vec n$ is the orphan spin, the only spin left in the cluster of the constrainer centred at $\textbf{R}_0$, and $m_{\rm O}$ is its constrainer coefficient.  A vacancy cluster is thus a constraint \emph{on the constrainer}; the Gauss-law gets modified such that Eq.~\eqref{eq:rankn-gauss-prl} acquires a charge density $\rho$ proportional to the orphan. To analyze the effect on the surroundings we use the hybrid field theory of Refs.~\cite{SenDamleMoessner12,FloresCalderonBentonMoessner24}, which treats the surrounding liquid in the self-consistent Gaussian approximation (SCGA) or equivalently large $\mathcal{N}$, with kernel $h_{ab}=\kappa\,\delta_{ab}+\beta J_{ab}$ and $\kappa$ a Lagrange multiplier enforcing the fixed spin-length on average, while keeping the orphan as an exact unit spin. The interaction matrix is $J_{ab}(\textbf{q})= J\,T^{a}(\textbf{q})T^{b}(\textbf{q})^*$. To orient the orphan we add a small uniform magnetic field, $-\sum_i\vec h\cdot\vec S_i$ with $\vec h=h\hat z$, to Eq.~\eqref{eq:constrainer-H-prl}.  It is weak enough to leave the liquid in its spin liquid regime, acquiring only a small uniform magnetization, linear in $h$, which we subtract, $\delta\langle\vec S_i\rangle=\langle\vec S_i\rangle-\langle\vec S_i\rangle_{{\rm ud},h}$, with $\rm ud$ denoting the undiluted system. To implement the orphan spin Eq.~\eqref{eq:orphan-is-charge-prl} we use a Lagrange multiplier $\vec\mu$, which adds $-\vec\mu\cdot(\vec C_0-m_{\rm O}\vec n)$ to the SCGA action.  Because $\vec C_0$ is linear in the spins, the orphan enters the liquid as a linear source $\vec\mu\cdot\vec C_0$ that couples through the constrainer itself.  Integrating out the spin liquid and imposing only the orphan spin constraint we obtain
\begin{equation}
\label{eq:texture-factorization-prl}
 \begin{aligned}
 \delta\langle\vec S^a_{\bR}\rangle&=\frac{\mathcal D^a(\bR)}{\sigma_0^2}\,\big(m_{\rm O}\langle\vec n\rangle-\langle\vec C_0\rangle_{{\rm ud},h}\big),\\
 \mathcal D^a(\bR)&=\frac{1}{N_{\rm uc}}\sum_{\bq}e^{\ii\bq\cdot(\bR+\bm\tau_a-\bR_0)}\,
 \frac{T^a(\bq)}{\kappa+\beta\,\omega(\bq)} ,
 \end{aligned}
\end{equation}
where $N_{\rm uc}$ is the number of unit cells, $\mathcal D^a(\bR)=\langle S^{a,z}_{\bR}C_0^z\rangle_{\rm ud}$ is the spin--charge correlator of the clean, undiluted liquid and $\sigma_0^2=\langle(C_0^z)^2\rangle_{\rm ud}$ the charge--charge correlator, and $\langle\vec n\rangle=\hat z\,L(\alpha h/T)$, with $L(x)=\coth x-1/x$, is the magnetization of a free spin carrying the renormalized--- rational or irrational \cite{FloresCalderonBentonMoessner24} ---orphan moment $\alpha=m_{\rm O}T\langle C^z_0\rangle_{{\rm ud},h}/h\sigma_0^2$. At low temperature the momentum-space constrainer is nonzero everywhere except in a neighborhood of the pinch-point that shrinks as $T\to0$, so that $\sigma_0^2\simeq T$. The spatial structure, and therefore winding, is supported by the undiluted system. The correspondence is thus a property of the spin liquid that a vacancy reveals.

We assume throughout that the entropic weight, quadratic in the low-energy fields is isotropic, i.e.\ invariant under rotations of the fields, as lattice symmetry enforces for all models studied here, the anisotropic case is discussed in the SM.  Near the band touching the Fourier transform of the texture is  $\delta\langle \Phi^z_A(\bk)\rangle=Q_{\rm eff}\,d_A(\bk)^*/[\kappa T+|\textbf d(\bk)|^2]$, with  $|\textbf d(\bk)|^2=\lambda k^{2n}$ and $\lambda$ a constant.  The spin texture is that of a point charge at short distances, neutralized by a screening cloud of radius $\xi_n=(\lambda/\kappa T)^{1/2n}$, the thermal length.  The orphan spin acts as a gauge charge of magnitude $Q_{\rm eff}=T\langle\mu^z\rangle=(T/\sigma_0^2)\,Q_{\rm core}$, where $Q_{\rm core}=m_{\rm O}\langle n^z\rangle-\langle C^z_0\rangle_{{\rm ud},h}$ is the charge imposed on the orphan constrainer.

\paragraph{Bulk--defect correspondence.---}
The topological invariant for the diluted system is a winding number now in real space. For a circle of radius $R$ around the orphan we have the spin texture $
 \hat{\bm\Phi}=\delta\langle\bm\Phi^z(\br)\rangle\big/|\delta\langle\bm\Phi^z(\br)\rangle| ,
$ with winding
\begin{equation}
 W_{\rm defect}=\frac{1}{2\pi}\oint_{|\br|=R}\dd x_\mu\;
 \epsilon_{AB}\,\hat\Phi_A\,\partial_{\mu}\hat\Phi_B \label{eq:defect-winding-prl}
\end{equation}
taken in the spin liquid window $a\ll R\ll\xi_n$. This defect winding number equals the bulk winding number under the following sufficient conditions: $\textbf d(\bk)$ is a homogeneous symbol of derivative order $n$ with maximal vorticity $|W_{\rm bulk}|=n$, as for a rank-$n$ spin liquid, non-vanishing in the proximity of the pinch point; the entropic weight of the low-energy fields is positive definite, as it is in the SCGA, and isotropic; and the orphan carries a nonzero gauge charge. Then we have
\begin{equation}
 W_{\rm defect}=W_{\rm bulk} \label{eq:rankn-theorem-prl}
\end{equation}
 A short proof is given in the End Matter; details are in the SM. It reduces to the two-dimensional Fourier transform acting \emph{diagonally} on angular harmonics: a momentum-space profile $e^{\ii m\phi}$ maps to a real-space profile $e^{\ii m\theta}$ times a radial factor that depends only on $|m|$. The spin texture has the long distance form $\delta\langle \Phi^z_c(\br)\rangle\propto Q_{\rm eff}\,r^{\,n-2}\, e^{\pm\ii n\theta}$ decaying as $1/r$ for rank 1, nondecaying at rank 2, and grows as $r^{n-2}$ for $n\ge3$, until it is cut off at the thermal length $\xi_n\propto T^{-1/2n}$.

\paragraph{Coulomb spin liquids: kagome and checkerboard lattices.---}

We proceed now to analyze in detail how the bulk-defect correspondence works in practice, first for the Coulomb spin liquids or rank-1 $\mathrm{U}(1)$ spin liquids. In this case since $n=1$ we have  $\textbf d^{(1)}(\bk)=\ii\,\mathcal M\bk$, with $\mathcal M$ a real nonsingular $2\times2$ matrix, so that $W_{\rm bulk}=\sgn\det\mathcal M$.  The gauge field is a linear combination of the soft modes $\vec B_\mu=\mathcal M_{A\mu}\vec\Phi_A$ and the projected constraint $d^{(1)}_A\vec\Phi_A=\ii k_\mu\vec B_\mu$ is the familiar Gauss law $\partial_\mu B_\mu=\rho$ for each spin component. In this case the field produced by the orphan spin is that of a magnetic monopole $\delta\langle\textbf B^z(\br)\rangle \rightarrow Q_{\rm eff}\hat{\br}/2\pi r$ in the large distance limit but smaller than the thermal length $r\ll\xi_1\propto T^{-1/2}$.

\begin{figure}[t]
\centering
\includegraphics[width=\columnwidth]{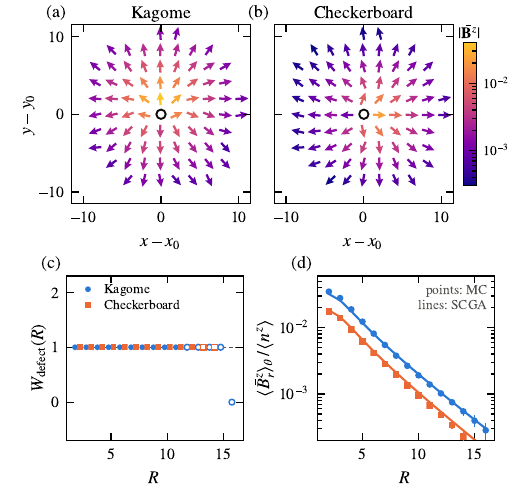}
\caption{Rank-1 bulk--defect correspondence in Monte Carlo ($L=48$, $T=0.05J$, $h=0.2J$).  (a,b)~Reconstructed flux $\bar{\textbf B}^z$ around the orphan (circle); color gives its magnitude.  (c)~Defect winding on circles of radius $R$; open symbols are not statistically robust.  (d)~Radial flux per unit $\langle n^z\rangle$, with the SCGA (lines).  Details in the SM.}
\label{fig:rank1-mc-winding}
\end{figure}
We provide two examples for the $n=1$ case, the Heisenberg model ($\mathcal{N}=3$) on the kagome [Fig.~\ref{fig:schematic}, top] and checkerboard lattices. Because they are line graphs, the spins sit on the links $\ell$ of a parent honeycomb (kagome) or square (checkerboard) lattice, and each constrainer is the total spin at a parent vertex $v$,
\begin{equation}
\label{eq:kc-hamiltonian-prl}
 \mathcal H_{\rm K,C}=\frac J2\sum_v\Big[\sum_{\ell\ni v}\vec S_\ell\Big]^2-\sum_\ell\vec h\cdot\vec S_\ell ,
\end{equation}
the constrainer form of Eq.~\eqref{eq:constrainer-H-prl}.  The checkerboard unit cell contains one parent vertex; the kagome cell contains two, and only the difference of their constrainers vanishes at the touching, so that it plays the role of $\textbf T(\bq)$ (SM).  Orienting every parent link from one sublattice of the bipartite parent lattice to the other makes the constrainer a lattice divergence, so the link spins \emph{are} the flux, $\mathcal M\propto\mathbb I$, and $W_{\rm bulk}=+1$.  The orphan keeps one spin at one parent vertex, so $m_{\rm O}=1$ and $Q_{\rm core}=\langle n^z\rangle-\langle C^z_0\rangle_{{\rm ud},h}$, and the rank-1 texture above predicts on both lattices a radial flux of winding one whose amplitude is fixed by the orphan magnetization.

We test this by classical Monte Carlo simulations of Eq.~\eqref{eq:kc-hamiltonian-prl} on both lattices in a uniform field $h=0.2J$, details of the implementation are discussed in the SM.  The undiluted spin liquid acquires a magnetization $m_h=0.034$ (kagome) and $0.024$ (checkerboard) per spin, which we measure in a clean reference run and subtract.  The orphan magnetization follows the free spin behaviour with $\alpha=0.517$ and $0.513$, close to the $T\to0$ moment prediction \cite{sen} of $\alpha=1/2$.  We reconstruct the flux $\bar{\textbf B}^z$ by coarse-graining the link magnetizations over a few lattice spacings and evaluate Eq.~\eqref{eq:defect-winding-prl} on circles around the orphan.  Figure~\ref{fig:rank1-mc-winding} shows the result.  On both lattices the flux is a hedgehog centered on the orphan [Fig.~\ref{fig:rank1-mc-winding}(a,b)], and its winding is exactly one on every circle from $R=2$ out to $R=11$ (kagome) and $R=12$ (checkerboard), more than three screening lengths [Fig.~\ref{fig:rank1-mc-winding}(c)], therefore $W_{\rm defect}=W_{\rm bulk}=1$. Indeed the thermal screening suppresses the magnitude of the field but not its winding. The radial envelope agrees well with the hybrid field theory texture of the diluted lattice in the same field Fig.~\ref{fig:rank1-mc-winding}(d). The plateau persists at $T=0.025J$ and $L=32$ and does not depend on the coarse-graining width, moreover a magnetic field applied to the orphan alone gives the same spin texture.

\begin{figure}[t]
\centering
\includegraphics[width=\columnwidth]{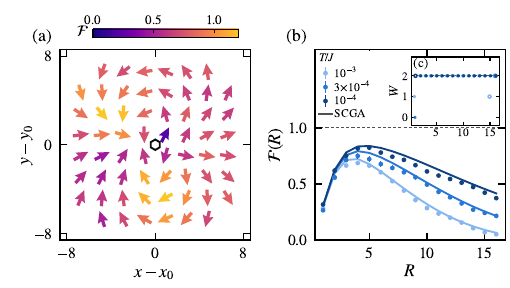}
\caption{Rank-2 bulk--defect correspondence in the honeycomb-snowflake model ($\gamma=1/2$, $L=48$, $h=0.2J$, $m_{\rm O}=1$).  (a)~Reconstructed $\Phi^z_c$ at $T=10^{-4}J$: arrows give its phase, color its magnitude $\mathcal F$ in units of its Coulomb-window value $Q_{\rm eff}/4\pi\sqrt\lambda$; the hexagon marks the orphan's snowflake.  (b)~Envelope $\mathcal F(R)$ at three temperatures, with the SCGA (lines).  Inset (c): defect winding of $\Phi_c$; open symbols are not statistically robust.  Details in the SM.}
\label{fig:rank2-mc-winding}
\end{figure}

\emph{Rank 2: the honeycomb-snowflake model.---}
The rank-2 $\mathrm{U}(1)$ spin liquid is realized microscopically in the honeycomb-snowflake model~\cite{BentonMoessner21,FloresCalderonBentonMoessner24},
\begin{equation}
\label{eq:rank2-hamiltonian-prl}
 \mathcal H_{\rm HS}=\frac J2\sum_{\hexagon} \Big[
 \sum_{i\in \hexagon}\vec S_i+\gamma\sum_{i\in\langle \hexagon\rangle}\vec S_i
 \Big]^2-\sum_i\vec h\cdot\vec S_i,
\end{equation}
where $\hexagon$ is a honeycomb hexagon and $\langle \hexagon \rangle$ denotes its six exterior neighbors, which together form the snowflake cluster of Fig.~\ref{fig:schematic} (bottom turwuoise links), the Hamiltonian follows the constrainer form of Eq.~\eqref{eq:constrainer-H-prl} with $N_c=1$ on $N=2$ sublattices.  At $\gamma=1/2$ the constrainer vanishes at the Brillouin-zone corners $\pm\bm K$~\cite{YanClassification}. Let us define the low-energy modes $\vec\psi_a=\vec S^a(\bm K+\bk)$ near $+\bm K$ on sublattice $a=1,2$, the expansion of Eq.~\eqref{eq:ground-constraint-prl} has orthonormal soft modes $\vec\Phi_1=(\vec\psi_1+\vec\psi_2)/\sqrt2$ and $\vec\Phi_2=\ii(\vec\psi_2-\vec\psi_1)/\sqrt2$ such that the expansion of the constrainer near $\bm K$ leads to $d^{(2)}(\bk)=\sqrt\lambda\,\big(k_x^2-k_y^2,\;2k_xk_y\big)$ with $\lambda=9/32$; since $N=2$ both soft modes are charged.  It follows then that $W_{\rm bulk}=2$ and the emergent gauge field is $B_{xx}=-B_{yy}=-\sqrt\lambda\,\Phi_1$ and $B_{xy}=-\sqrt\lambda\,\Phi_2$, obeying $\partial_\mu\partial_\nu B_{\mu\nu}=\rho$ in each spin component. The vacancy clusters leave a single spin in one snowflake, with $m_{\rm O}=1$ or $\gamma$ according to whether it sits on the hexagon or outside it. The gauge field surrounding the orphan spin tends to $B_c=B_{xx}+\ii B_{xy}=(Q_{\rm eff}/4\pi)\,e^{2\ii\theta}$, so that $\Phi_c=-B_c/\sqrt\lambda$, for large distances and $r\ll\xi_2=(\lambda/\kappa T)^{1/4}$, the field is non-decaying for a rank-2 spin liquid as shown in Ref.~\cite{FloresCalderonBentonMoessner24} and winds two times over the origin therefore $W_{\rm defect}=W_{\rm bulk}=2$.

We provide numerical evidence of the winding with Monte Carlo simulations as for the rank-1 case here at $\gamma=1/2$  for $L=48$, $T=10^{-4}J$ in the uniform field $h=0.2J$, with a hexagon orphan ($m_{\rm O}=1$).
The orphan follows a free spin with magnetic moment $\alpha=0.226(2)$, close to the $T\to0$ moment $2/9$.
The reconstructed field turns twice around the orphan [Fig.~\ref{fig:rank2-mc-winding}(a)], and its winding is $W_{\rm defect}=2$ on every circle from $R=2$ to $R=16$, about $2.4\,\xi_2$. The enclosed valley charge  $Q_{\rm eff}=0.944(1)$ is consistent with the SCGA value as well as the screening length $T^{-0.26(1)}$ with the predicted $T^{-1/4}$. Similar results, with the same winding invariant, hold for the exterior orphan with $m_{\rm O}=\gamma$.

\paragraph{Discussion.---}
We establish a bulk--defect correspondence between the momentum-space winding of an isolated pinch-point and the real-space winding of a vacancy-induced spin texture. The gauge charge character of the orphan spin allows for long-range fields with the undiluted spin liquid determining its spatial structure. We present a unified proof for rank-1, rank-2, and higher-rank spin liquids. Additionally we confirm our results with Monte Carlo simulations of the kagome, checkerboard, and honeycomb-snowflake Heisenberg models.

A vacancy thus provides a local probe of the pinch-point singularity, the winding of the spin texture may be used generically to distinguish higher-rank spin liquids from conventional Coulomb phases. Diluted kagome-based magnets, where NMR resolves defect-induced textures~\cite{Limot02,Hagemann01,sen}, offer a possible route to explore this prediction. Other possible platforms are artificial square ice~\cite{Perrin16,King21} and frustrated honeycomb magnets~\cite{Rehn16}.
We further comment that since the phenomenology of orphan spins survives for quantum systems at finite temperatures above the ordering scale and below the exchange interaction energy we expect our results may apply for large $S$-spin models and possibly even spin $1/2$ systems ~\cite{Huang16,Patil20,Schafer20}. For authentic quantum spin liquids the full dynamical gauge theory is needed and is an interesting future direction. Beyond exploring the quantum models we conclude by mentioning that understanding the topological invariant and correspondence for higher-rank classical spin liquids in three dimensions remains an open question.

\begin{acknowledgments}
\paragraph{Acknowledgments.---}
The author gratefully acknowledges insightful discussions with Roderich Moessner and Owen Benton.
\end{acknowledgments}

\bibliography{references}

\section*{End Matter}

\paragraph{Proof of the bulk--defect correspondence.---}
We prove Eq.~\eqref{eq:rankn-theorem-prl} for the texture of Eq.~\eqref{eq:texture-factorization-prl}, in the hybrid SCGA with the orphan multiplier alone and in the continuum limit near the touching (SM).  The proof rests on one property of the plane: the Fourier transform commutes with rotations, so it cannot change the angular harmonic of a field.  The conditions below make the symbol, and hence the texture, a single harmonic.

We write two-component objects as complex numbers, $d_c=d_1+\ii d_2$ and $\Phi_c=\Phi_1+\ii\Phi_2$, with $\zeta=k_x+\ii k_y=ke^{\ii\phi}$, $u=e^{\ii\phi}$, and $re^{\ii\theta}$ in real space.  Since $\epsilon_{AB}\hat f_A\,\dd\hat f_B=\dd\arg f_c$, Eqs.~\eqref{eq:general-framework-prl} and \eqref{eq:defect-winding-prl} are the windings $\deg f=\frac1{2\pi}\oint\dd\arg f$ of $\ii^{-n}d_c$ around $\bq_0$ and of $\delta\langle\Phi^z_c\rangle$ around the orphan.
 We assume
\begin{enumerate}[label=(H\arabic*),leftmargin=2.3em,itemsep=0pt,topsep=2pt]
 \item a two-component elliptic symbol: the charged sector is the real two-plane fixed by $\Theta$, on which $\ii^{-n}\textbf d(\bk)$ is a real homogeneous polynomial of derivative order $n$, nonzero for $\bk\neq0$, so that $W_{\rm bulk}$ is defined;
 \item maximal vorticity, $|W_{\rm bulk}|=n$;
 \item a positive and isotropic Gaussian weight: the stiffness of the charged sector is rotation invariant, $|\textbf d(\bk)|^2=\lambda k^{2n}$;
 \item a nonzero gauge charge, $Q_{\rm eff}\neq0$, and an invertible map from the measured spins to the charged field.
\end{enumerate}

\emph{Proof.--}  A homogeneous polynomial of degree $n$ in $(k_x,k_y)$ is a combination of the monomials $\zeta^j\bar\zeta^{\,n-j}=k^ne^{\ii(2j-n)\phi}$, $j=0,\dots,n$, so by (H1), with $P(w)=\sum_{j=0}^np_jw^j$,
\begin{equation}
\label{eq:em-polynomial}
 \ii^{-n}d_c(\bk)=\sum_{j=0}^np_j\,\zeta^j\bar\zeta^{\,n-j}=k^nu^{-n}P(u^2).
\end{equation}
Its angular harmonics are $m=2j-n\in\{-n,-n+2,\dots,n\}$.  As $\phi$ runs once around, $u^{-n}$ winds $-n$ times and $w=u^2$ runs twice around the unit circle, so the argument principle gives $W_{\rm bulk}=2Z(P)-n$, with $Z(P)$ the number of zeros of $P$ in $|w|<1$; ellipticity keeps them off the circle.  Maximal vorticity alone requires only that all $n$ zeros lie inside the disc, or none.  Isotropy is stronger.  It means $|P(w)|^2=\lambda$ on $|w|=1$, and since there $\overline{P(w)}=\bar P(1/w)$, with $\bar P$ the polynomial of conjugated coefficients, the identity $P(w)\,w^n\bar P(1/w)=\lambda w^n$ holds for all $w$.  Both factors on the left are polynomials, so every zero of $P$ lies at $w=0$, and $P=p_jw^j$ is a monomial of winding $2j-n$.  With (H2) only $j=n$ or $j=0$ remain:
\begin{equation}
\label{eq:em-symbol}
 \ii^{-n}d_c(\bk)=\sqrt\lambda\,e^{\ii\chi}\,k^n u^{\pm n},
\end{equation}
with a constant phase $\chi$.  The symbol is the single harmonic $m=\pm n$.  It is then also harmonic as a polynomial, so the constant matrix $L$ of the main text exists, with $LL^T=\lambda\mathbb I$, and $\textbf B=L\bm\Phi$ is the symmetric traceless rank-$n$ gauge field.

The orphan enters the Gaussian action as a linear source through its own constrainer, and Eq.~\eqref{eq:texture-factorization-prl} is the response of the liquid to that source.  Projected onto the charged soft modes near the touching it reads $
 \delta\langle\Phi^z_A(\bk)\rangle=Q_{\rm eff}\,\frac{d_A(\bk)^*}{\kappa T+|\textbf d(\bk)|^2}.$
The numerator is the constrainer through which the orphan couples and the denominator the stiffness of the dispersive band; the induced charge density $\textbf d\cdot\delta\langle\bm\Phi^z\rangle$ is a point charge $Q_{\rm eff}$ for $k\gg\xi_n^{-1}$, with $\xi_n=(\lambda/\kappa T)^{1/2n}$, screened for $k\lesssim\xi_n^{-1}$.  Since $\ii^{-n}\textbf d$ is real, $d_A^*=(-\ii)^n\,\ii^{-n}d_A$, and the texture with the isotropic denominator gives
\begin{equation}
\label{eq:em-response}
 \delta\langle\Phi^z_c(\bk)\rangle=Q_{\rm eff}\sqrt\lambda\,e^{\ii\chi}\,(-\ii)^n\,
 \frac{k^n}{\kappa T+\lambda k^{2n}}\,e^{\pm\ii n\phi}.
\end{equation}
Because the denominator depends on $k$ only, temperature changes the radial profile but not the angular dependence: the response is the single harmonic $m=\pm n$ at every $T$.

A rotation of the plane by $\alpha$ multiplies a harmonic $e^{\ii m\phi}$ by $e^{-\ii m\alpha}$, and the Fourier transform commutes with rotations.  It therefore maps each harmonic to the same harmonic in real space and acts only on the radial profile.  Explicitly, inserting the Jacobi--Anger expansion $e^{\ii\bk\cdot\br}=\sum_l\ii^lJ_l(kr)e^{\ii l(\phi-\theta)}$, the angular integral selects $l=-m$, and with $J_{-m}=(-1)^mJ_m$ and $\int_{\bk}\equiv\int\dd^2k/(2\pi)^2$
\begin{equation}
\label{eq:em-hankel}
 \int_{\bk}e^{\ii\bk\cdot\br}f(k)\,e^{\ii m\phi}
 =\frac{\ii^{\,m}e^{\ii m\theta}}{2\pi}\int_0^\infty\!\dd k\,k\,f(k)J_m(kr)
\end{equation}
for any radial profile $f(k)$: a Hankel transform of order $m$.  For Eq.~\eqref{eq:em-response}, $m=\pm n$ and the phases combine to $(-\ii)^n\ii^{\pm n}J_{\pm n}=J_n$.  Substituting $s=kr$,
\begin{equation}
\label{eq:em-texture}
 \begin{aligned}
 \delta\langle\Phi^z_c(\br)\rangle&=\frac{Q_{\rm eff}\,e^{\ii\chi}}{2^n\pi\,(n-1)!\sqrt\lambda}\,r^{\,n-2}e^{\pm\ii n\theta}\,F_n\!\Big(\frac r{\xi_n}\Big),\\
 F_n(x)&=2^{n-1}(n-1)!\int_0^\infty\!\dd s\,\frac{s^{n+1}J_n(s)}{s^{2n}+x^{2n}} .
 \end{aligned}
\end{equation}
The envelope is real and tends to 1 for $x\to0$, by $\int_0^\infty\dd s\,s^{1-n}J_n(s)=2^{1-n}/(n-1)!$~\cite{SteinWeiss}; $F_1(x)=xK_1(x)$ is the screened Coulomb monopole.  On a circle of radius $R$ the phase of Eq.~\eqref{eq:em-texture} is $\chi\pm n\theta$, shifted by $\pi$ where $F_n<0$, so it advances by $\pm2\pi n$: $W_{\rm defect}=\pm n$ on every circle on which $F_n(R/\xi_n)\neq0$, in particular throughout the window $a\ll R\ll\xi_n$, where $F_n\simeq1$.  The sign is that of $u^{\pm n}$ in Eq.~\eqref{eq:em-symbol}, so $W_{\rm defect}=W_{\rm bulk}$. $\square$

\emph{ Independence of the basis and anisotropy}  The measured field is related to the soft modes by constant invertible maps, the reconstruction from the spins and $\textbf B=L\bm\Phi$.  Under $\bm\Phi\to M\bm\Phi$ the constraint $\textbf d\cdot\bm\Phi$ is kept by $\textbf d\to M^{-T}\textbf d$, and both windings are multiplied by $\sgn\det M=\sgn\det M^{-T}$, so the equality holds in every basis.   An anisotropic stiffness would populate the \emph{infrared channels} $|m|\le n-2$, whose Hankel integrals diverge at small $k$, since $J_m(s)\sim s^{|m|}$; cut off at $\xi_n^{-1}$, they are enhanced in the window by $(\xi_n/r)^{n-2-|m|}$ and can change the winding.  The SM shows that the correspondence survives for the unscreened saddle at any $n$ and at finite temperature for $n\le2$, and lattice rotations enforce (H3), as for the three models studied here.

\clearpage
\onecolumngrid
\begin{center}
{\large\bfseries Supplemental material for ``Bulk--Defect Correspondence in Algebraic Classical Spin Liquids''}\\[4mm]
Rafael Flores-Calder\'on$^{1,2,*}$\\[2mm]
{\small\itshape $^1$Technical University of Munich, TUM School of Natural Sciences, Physics Department, 85748 Garching, Germany\\
$^2$Munich Center for Quantum Science and Technology (MCQST), Schellingstr.~4, 80799 M\"unchen, Germany\\
$^*$Electronic address: r.flores-calderon@tum.de}\\[2mm]
{\small (Dated: \today)}
\end{center}
\vspace{2mm}
\setcounter{equation}{0}\renewcommand{\theequation}{S\arabic{equation}}
\setcounter{figure}{0}\renewcommand{\thefigure}{S\arabic{figure}}
\setcounter{table}{0}\renewcommand{\thetable}{S\arabic{table}}

\makeatletter
\newcommand{\smtocline}[2]{\noindent\hyperref[sec:S#1]{\makebox[3em][l]{\bfseries S#1.}{\bfseries #2}\nobreak\leaders\hbox{$\m@th\mkern2mu\hbox{.}\mkern2mu$}\hfill\nobreak\hbox to 1.6em{\hfil\bfseries\pageref*{sec:S#1}}}\par}
\makeatother
\begin{center}
{\color{RubineRed}\bfseries Contents}
\end{center}
\vspace{-1mm}
\smtocline{1}{Constrainers and their real structure}
\smtocline{2}{Vacancies and gauge charges: the hybrid field theory}
\smtocline{3}{Topological invariants and the gauge field}
\smtocline{4}{The rank-$n$ bulk--defect correspondence: proof and conditions}
\smtocline{5}{Rank 1: Coulomb spin liquids}
\smtocline{6}{Rank 2: tensor gauge field}
\smtocline{7}{Microscopic models}
\smtocline{8}{Monte Carlo simulations}
\bigskip

\emph{Notation.}  We use the conventions of the main text throughout: unit cells $\bR$ ($N_{\rm uc}$ of them, of area $A_{\rm uc}$), sublattices $a,b=1,\dots,N$, and constrainers $l=1,\dots,N_c$, whose Fourier components $T^a_l(\bq)\equiv\mathsf T_{la}(\bq)$ form the $N_c\times N$ matrix $\mathsf T$ with rows $\textbf T_l$; energies are in units of $J$, $\bq_0$ is the band touching, and $\bk=\bq-\bq_0$.  The charged soft modes $\vec\Phi_A$, $A,B=1,2$, the symbol $\textbf d(\bk)$, the stiffness $\lambda$, defined by $|\textbf d(\bk)|^2=\lambda k^{2n}$ for the isotropic Gaussian weight assumed throughout [condition (H3) of Sec.~S4], the gauge field $B$ with independent components $B_1=B_{xx\cdots x}$ and $B_2=B_{xx\cdots xy}$, the canonical symbol $\textbf d_0(\bk)=k^n(\cos n\phi,\sin n\phi)$, and the matrix $L$ in $B_A=L_{AB}\Phi_B$ are defined in the main text for one constrainer per unit cell and in Sec.~S1 for $N_c$ constrainers.  Where components are needed we write $\Phi_{AI}$, with $I,J$ collecting spin components and any other spectator labels (valleys, symmetry-equivalent copies); $\mu,\nu$ are spatial directions.  At rank 1, $\textbf d=\ii\mathcal M\bk$ and $B_{\mu I}=\mathcal M_{A\mu}\Phi_{AI}$.  Sections~S1--S4 develop the general rank-$n$ theory: the real structure of the constrainer, the vacancy as a gauge charge, the invariants and the gauge field, and the bulk--defect theorem with its conditions.  Sections~S5 and S6 specialize it to rank 1 and rank 2, Sec.~S7 gives the microscopic models, and Sec.~S8 the Monte Carlo simulations of all three lattices.

\subsection*{S1. Constrainers and their real structure}\phantomsection\label{sec:S1}

We first make precise the connection to the constrainer formalism of Ref.~\cite{YanClassification}.  A linear constrainer and its Fourier transform may be written as
\begin{equation}
\label{eq:supp-constrainer-def}
 \vec C_l(\bR)=\sum_{\br,a}c_{la}(\br)\vec S^a(\bR+\br),
 \qquad
 \vec C_l(\bq)=\sum_aT^a_l(\bq)^*\vec S^a(\bq),
\end{equation}
with $c_{la}(\br)\equiv C^a_l(0,\br)$ and $T^a_l(\bq)=\mathsf T_{la}(\bq)=\sum_{\br}c_{la}(\br)e^{-\ii\bq\cdot\br}$; the spin components are spectator labels and are suppressed here.  We use position-based phases: $\br$ is the position of the site relative to the centre of the constrainer cluster.  A cell-based convention multiplies $\mathsf T$ by $\bq$-dependent diagonal phases on sites and constrainers (sewing matrices); the symmetry relations below then carry these phases, while the leading effective symbol changes only by a constant unitary.  For $N_c$ constrainers per unit cell the Hamiltonian is $\mathcal H=\frac J2\sum_{\bR,l}\vec C_l(\bR)^2$, the interaction matrix is $J_{ab}=J\sum_lT^a_lT^{b\,*}_l$, and the ground state obeys $\vec C_l(\bq)=0$ for every $l$ and $\bq$.  If the rows $\textbf T_l(\bq)$ are linearly independent away from the touching, there are $N-N_c$ flat bands (for $N>N_c$) and $N_c$ dispersive bands, the eigenvalues $g_b(\bq)$ of the Gram matrix $G_{ll'}=\textbf T_l^\dagger\textbf T_{l'}$ (Sec.~S2).  A pinch point is a momentum $\bq_0$ at which the rank of $\mathsf T$ drops to $r_0<N_c$, and we assume a single low-energy sector, $N_c-r_0=1$.  Let $U_0$ and $V_0$ be orthonormal bases of the left and right null spaces of $\mathsf T(\bq_0)$ and $U_1$, $V_1$ their complements, and write $\mathsf T(\bq_0+\bk)$ in these bases as the blocks $A=U_1^\dagger\mathsf TV_1$, invertible near $\bq_0$, $B=U_1^\dagger\mathsf TV_0$, $C=U_0^\dagger\mathsf TV_1$ and $D=U_0^\dagger\mathsf TV_0$.  Eliminating the gapped modes from $\mathsf T\vec S=0$ leaves the low-energy constraint $D_{\rm eff}=D-CA^{-1}B$, a $(N_c-r_0)\times(N-r_0)$ matrix whose single row in the rank-$n$ sector is the symbol $\textbf d(\bk)$.  The soft modes $\vec\Phi_A=\sum_a\mathcal U_{Aa}\vec S^a(\bq_0+\bk)$, $A=1,\dots,N-r_0$, are the coordinates in the orthonormal basis $\mathcal U$ of $V_0$ in which $D_{\rm eff}$ is written, so that the constraint reads $d_A\vec\Phi_A=0$ at leading order; the charged sector is the span of $\textbf d$, $A=1,2$, and the remaining $N-r_0-2$ modes are neutral.  For $N_c=1$, as in the main text, $r_0=0$, every mode is soft at $\bq_0$, there is no gapped block, and $T^a(\bq_0+\bk)^*=d_A(\bk)\,\mathcal U_{Aa}+O(k^{n+1})$.  Since $B,C=O(k)$, the direct projection $D=U_0^\dagger\mathsf TV_0$ suffices at $n=1$, but for $n\ge2$ the correction $CA^{-1}B$, which starts at $O(k^2)$, enters at or below the order of $\textbf d$; in the honeycomb-snowflake and checkerboard models $\mathsf T(\bq_0)=0$, so there is no gapped block.

\paragraph*{Real structure from inversion and reality.}
Both ingredients of $\Theta$ act directly on Eq.~\eqref{eq:supp-constrainer-def}.  Since the coefficients $c_{la}(\br)$ are real numbers,
\begin{equation}
\label{eq:supp-reality}
 \mathsf T_{la}(\bq)^{*}=\sum_{\br}c_{la}(\br)\,e^{+\ii\bq\cdot\br}=\mathsf T_{la}(-\bq).
\end{equation}
Let $P$ be inversion about a centre taken as the phase origin, so that it maps $\br\mapsto-\br$.  It carries the site $\bR+\bm\tau_a$ to $-\bR-\bm\tau_a$, which is again a site: inversion symmetry of the lattice is precisely the statement that for every $a$ there is a sublattice $\bar a$ with
\begin{equation}
\label{eq:supp-bar-a}
 \bm\tau_{\bar a}=-\bm\tau_a \pmod{\text{Bravais lattice}},
\end{equation}
and $a\mapsto\bar a$ is then an involution, $\bar{\bar a}=a$.  Writing $\bar l$ for the image of the cluster of constrainer $l$, invariance of the Hamiltonian reads $c_{\bar l\bar a}(-\br)=c_{la}(\br)$.  Relabelling the sum by $\br\mapsto-\br$,
\begin{equation}
\label{eq:supp-inversion}
 \mathsf T_{\bar l\bar a}(\bq)=\sum_{\br}c_{\bar l\bar a}(-\br)\,e^{+\ii\bq\cdot\br}
 =\sum_{\br}c_{la}(\br)\,e^{+\ii\bq\cdot\br}=\mathsf T_{la}(-\bq).
\end{equation}
Equations~\eqref{eq:supp-reality} and \eqref{eq:supp-inversion} have the same right-hand side, so composing them removes $-\bq$ altogether and leaves a constraint at fixed momentum,
\begin{equation}
\label{eq:supp-theta-symbol}
 \mathsf T_{la}(\bq)^{*}=\mathsf T_{\bar l\,\bar a}(\bq),
\end{equation}
valid at every $\bq$, and in particular everywhere on the loop $\mathcal C$.  This is the symmetry $\Theta=P\mathcal K$; it is an involution, $\Theta^{2}=P^{2}\mathcal K^{2}=\mathbb I$, because $P$ permutes sublattices and therefore has real matrix elements.  The set of $\Theta$-fixed vectors is a \emph{real} subspace and not a complex one, since $\Theta(\ii v)=-\ii\Theta v$.  $\Theta^2=\mathbb{I}$ makes it a real form of every $\Theta$-invariant complex subspace $V$: for any $v\in V$,
\begin{equation}
\label{eq:supp-real-form}
 v_{+}=\tfrac12\left(v+\Theta v\right),
 \qquad
 v_{-}=\tfrac{1}{2\ii}\left(v-\Theta v\right)
\end{equation}
are both $\Theta$ fixed---using $\Theta^{2}=\mathbb{I}$, and for $v_{-}$ also that antilinearity conjugates the prefactor---while $v=v_{+}+\ii v_{-}$.  Every vector of $V$ therefore splits uniquely into a fixed part and $\ii$ times a fixed part, so a real basis of the fixed part of $V$ is a complex basis of $V$, of the same dimension.  $\Theta^2=\mathbb I$ thus provides the real structure but not the dimension.  The charged sector, the span of the symbols $\textbf d(\bk)$, is $\Theta$ invariant by Eq.~\eqref{eq:supp-theta-symbol}; that it is a complex plane, so that its fixed set is a real two-plane with basis $\{f_1,f_2\}$, is the hypothesis of a single isolated rank-$n$ sector with a two-component symbol [(H1) of Sec.~S4].  It is automatic at $n=1$, where $\textbf d$ is linear in $\bk$, but not for $n\ge2$, where a degree-$n$ symbol can span up to $n+1$ dimensions.  Had $\Theta$ squared to $-\mathbb{I}$, $\Theta v=v$ would force $v=-v$ and the fixed set would be trivial; the particular invariant used here, the degree of a real circle map, would then be unavailable, although other symmetry-protected invariants are not excluded.  Projecting Eq.~\eqref{eq:supp-theta-symbol} onto the two active soft modes therefore yields, in the basis $\{f_1,f_2\}$---where $\Theta$ acts as componentwise complex conjugation---a real projected symbol, up to the overall $\ii^{\,n}$ discussed in Sec.~S4; equivalently, in the rotated basis $e_{1,2}=(f_1\mp\ii f_2)/\sqrt2$, for which $\Theta e_1=e_2$, the conjugate-pair form
\begin{equation}
\label{eq:supp-conjugate-pair}
 d_1(\bk)=d_2(\bk)^{*} .
\end{equation}
The two are related by a fixed unitary and carry the same content; writing $d_1=\varrho e^{\ii\varphi}$ with $\varrho>0$, the pair is fixed by the single real two-vector $\textbf d=(\operatorname{Re}d_1,\operatorname{Im}d_1)$ whose normalization is the circle map $\hat{\textbf d}$ of the main text.

Which of the two forms is the natural one depends on the lattice, because $\bar a$ need not be a nontrivial permutation.  On the honeycomb, inversion about a bond midpoint or a hexagon centre gives $\bm\tau_{\bar 1}=-\bm\tau_1$ with $\bar 1=2$: the sublattices are exchanged, the snowflake constrainer maps to itself, and Eq.~\eqref{eq:supp-theta-symbol} reads $\mathsf T_1(\bq)^{*}=\mathsf T_2(\bq)$, which is the origin of the conjugate pair $(k_+^2,k_-^2)$ in Eq.~\eqref{eq:supp-rank2-constrainer-expansion}.  On kagome, with the inversion centre at a site, Eq.~\eqref{eq:supp-bar-a} is solved by $\bar a=a$: the sublattices are \emph{not} exchanged, and it is instead the two simplex constrainers that are, $\bar l\ne l$.  The singular row is their difference, in position-based phases $\bm C_{\rm K}(\bq)=-2\ii\big(\sin\tfrac{\bq\cdot\hat{\bm e}_a}{2}\big)_{a=0,1,2}$, and Eq.~\eqref{eq:supp-theta-symbol} gives $\bm C_{\rm K}(\bq)^{*}=-\bm C_{\rm K}(\bq)$ at every $\bq$: the row is purely imaginary, which is the real form with $n=1$.  The cell-based row of Eq.~\eqref{eq:supp-kagome-small-q-row} differs from it by the site phases $e^{\ii\bq\cdot\hat{\bm e}_a/2}$ and the phase $e^{\ii\bq\cdot\hat{\bm e}_0}$ of the $\mathcal B$ simplex, at linear order by $\ii(\bq\cdot\hat{\bm e}_0)(1,1,1)$, which lies along the gapped constraint and drops out of the soft projection, so that the two give the same leading symbol.  On checkerboard, with the inversion centre at a parent vertex, the sublattices and the single constrainer of the primitive cell of Sec.~S7 are all mapped to themselves.  In position-based phases its row, $2(\cos\frac{q_x}2,\cos\frac{q_y}2)$, is real at every $\bq$, and the staggered variables of Eq.~\eqref{eq:supp-checkerboard-flux-reconstruction} turn it into $2\ii(\sin\frac{k_x}2,\sin\frac{k_y}2)$, the purely imaginary form of Eq.~\eqref{eq:supp-checkerboard-principal-symbol}.  Moving the phase origin multiplies $\mathsf T$ by a common $\bq$-dependent phase; this does not affect the degree, only the orientation convention fixed in Sec.~S7.

\subsection*{S2. Vacancies and gauge charges: the hybrid field theory}\phantomsection\label{sec:S2}

This section derives Eq.~\eqref{eq:texture-factorization-prl} of the main text, the screened monopole that follows from it, and the identification of the gauge charge, before any rank is specified.  We use the hybrid field theory of Ref.~\cite{FloresCalderonBentonMoessner24}, in which the orphan is kept as an exact unit spin while the surrounding liquid is treated in the self-consistent Gaussian approximation (SCGA).

\paragraph*{Self-consistent Gaussian approximation and vacancy clusters.}
Relax the unit length of the spins and impose it on average through a multiplier $\kappa$ fixed by $\langle\vec S_i^{\,2}\rangle=1$ (the quantity called $\rho$ in Ref.~\cite{FloresCalderonBentonMoessner24}).  With $\vec S^a_{\bR}=N_{\rm uc}^{-1/2}\sum_{\bq}e^{\ii\bq\cdot(\bR+\bm\tau_a)}\vec S^a(\bq)$ on $N_{\rm uc}$ unit cells, the weight is $e^{-\mathcal S_0}$ with
\begin{equation}
\label{eq:supp-hybrid-clean-action}
 \mathcal S_0=\frac12\sum_{\bq,ab}
 \vec S^{a}(\bq)^*\,h_{ab}(\bq)\,\vec S^{b}(\bq)
 -\beta\,\vec h\cdot\sum_{\bR,a}\vec S^a_{\bR},
 \qquad
 h_{ab}(\bq)=\kappa\,\delta_{ab}
 +\beta\,J_{ab}(\bq),
\end{equation}
and $J_{ab}=\sum_lT^a_lT^{b\,*}_l$ the interaction matrix, $T^aT^{b\,*}$ of the main text for $N_c=1$.  The same constrainer that locates the pinch point therefore also fixes the Gaussian weight.  We write $\langle\cdots\rangle_{\rm ud}$ for averages in the undiluted liquid and $\langle XY\rangle_{\rm ud}$ for connected correlations per spin component.

In the diluted problem the vacancy sites carry no spin, while the orphan keeps unit length, $\vec S_{\rm O}=\vec n$ with $\vec n^{\,2}=1$.  For the clusters used here the constrainer $l_0$ of the cluster centred at $\bR_0$ collapses onto the surviving spin, Eq.~\eqref{eq:orphan-is-charge-prl},
\begin{equation}
\label{eq:supp-orphan-constraint}
 \vec C_0=m_{\rm O}\,\vec n,
 \qquad
 m_{\rm O}=\begin{cases}
 1,&\text{kagome, checkerboard, hexagon site of a snowflake},\\
 \gamma,&\text{exterior site of a snowflake},
 \end{cases}
\end{equation}
with the clusters of Sec.~S7.  The hybrid partition function is
\begin{equation}
\label{eq:supp-hybrid-partition}
 Z=\int\mathcal D\vec S\int\dd\vec n\;\delta(\vec n^{\,2}-1)\,
 \prod_{v}\delta^{(\mathcal{N})}\!\left(\vec S_{v}\right)\,
 \delta^{(\mathcal{N})}\!\left(\vec C_0-m_{\rm O}\vec n\right)
 e^{-\mathcal S_0+\beta\vec h_{\rm O}\cdot\vec n} .
\end{equation}
Here $\mathcal S_0$ contains the uniform field $\vec h$ acting on the liquid, Eq.~\eqref{eq:supp-hybrid-clean-action}, and $\vec h_{\rm O}$ is a field on the orphan alone, used for the cross-check of Sec.~S8.  In the uniform-field protocol $\vec h_{\rm O}=0$: the vacancy factors pin the other spins of the cluster to zero, so Eq.~\eqref{eq:supp-orphan-constraint} makes the liquid spin at the orphan site equal to $\vec n$, and its Zeeman energy is already in $\mathcal S_0$; a separate term $\beta\vec h\cdot\vec n$ would count the orphan moment twice.
The constrainer is the lattice charge density: it vanishes in the bulk ground-state manifold, and the vacancy cluster pins it to $m_{\rm O}\vec n$ at a single site of the dual lattice, so no charge is imposed by hand.  The vacancy factors $\delta(\vec S_v)$ are exponentiated like the orphan factor below, with one multiplier per vacancy site. From this starting point several approximations follow that simplify the problem.

 (a)~\emph{All constraints.} First we can take all vacancy constraints and orphan moment into account. For one spin component write the spins as a vector $\textbf s$, the Gaussian kernel of Eq.~\eqref{eq:supp-hybrid-clean-action} in real space as $M$, the field source as $\textbf f=\beta h\textbf 1$, and the constraints as $E^T\textbf s=\textbf g$, where the columns of $E$ are the unit vectors of the vacancy sites and the coefficient vector $\textbf c_0$ of the orphan constrainer, and $\textbf g=(0,\dots,0,m_{\rm O}n^z)$.  Stationarity of $\frac12\textbf s^TM\textbf s-\textbf f^T\textbf s-\bm\mu^T(E^T\textbf s-\textbf g)$ gives
\begin{equation}
\label{eq:supp-full-multipliers}
 M\textbf s=\textbf f+E\bm\mu,\quad E^T\textbf s=\textbf g
 \quad\Rightarrow\quad
 \bm\mu=\big(E^TM^{-1}E\big)^{-1}\big(\textbf g-E^TM^{-1}\textbf f\big),\qquad
 \delta\textbf s=M^{-1}E\bm\mu ,
\end{equation}
with $\delta\textbf s$ the texture relative to the undiluted solution $M^{-1}\textbf f$; Eq.~\eqref{eq:supp-lattice-saddle} of Sec.~S8 is its lattice evaluation.

 (b)~\emph{One source.} A further approximation is to realize that the charged object will give the main long-range effect so that one can drop the vacancy delta funtions and focus on the orphan spin as one souce. Keeping only the orphan column, $E\to\textbf c_0$, gives $E^TM^{-1}E=\sigma_0^2$, $E^TM^{-1}\textbf f=\langle C^z_0\rangle_{{\rm ud},h}$ and $\delta\textbf s=\mathcal D\mu_0$, which is Eqs.~\eqref{eq:supp-hybrid-mean}--\eqref{eq:supp-hybrid-mu-saddle} below; it is exact when the one-source solution already satisfies the vacancy constraints.  The vacancy columns share the cluster sites with $\textbf c_0$, so once they are kept $\mu_0$ alone is no longer the monopole amplitude: at $T=0.05J$ they raise $T\mu_0$ by $8\%$ on kagome and $10\%$ on checkerboard (by less than $0.4\%$ at rank 2), yet on the lattices of Sec.~S7 they change the enclosed charge of Sec.~S8 by less than $2\times10^{-4}$ for $R\ge1$, and the coarse-grained rank-1 flux by less than $3\%$ for $R\ge4$.  With vacancies kept we therefore quote the amplitude of the lattice solution, $Q_{\rm eff}^{\rm SCGA}$ of Sec.~S8.

 (c)~\emph{Continuum.}  Finally one can use the low-energy expressions ear an isolated touching, then the propagator $\mathcal D$ is replaced by its pinch-point form which arises from the emergent gauge theory description.

Impose the orphan delta function with a Lagrange multiplier $\vec\mu$, integrated along the imaginary axis,
\begin{equation}
\label{eq:supp-hybrid-multiplier}
 \delta^{(\mathcal{N})}\!\left(\vec C_0-m_{\rm O}\vec n\right)
 =\int_{-\ii\infty}^{\ii\infty}\!\frac{\dd^{\mathcal{N}}\mu}{(2\pi\ii)^{\mathcal{N}}}\,
 e^{\vec\mu\cdot(\vec C_0-m_{\rm O}\vec n)},
 \qquad
 \vec C_0=\frac{1}{\sqrt{N_{\rm uc}}}\sum_{\bq,a}e^{\ii\bq\cdot\bR_0}\,T^a_{l_0}(\bq)^*\,\vec S^a(\bq).
\end{equation}
The hybrid weight is then $e^{-\mathcal S}$ with the action
\begin{equation}
\label{eq:supp-hybrid-action}
 \mathcal S[\vec S,\vec\mu]=\mathcal S_0[\vec S]-\vec\mu\cdot\left(\vec C_0-m_{\rm O}\vec n\right)-\beta\vec h_{\rm O}\cdot\vec n ,
\end{equation}
in which the spin length is held on average by $\kappa$ and the vacancy constraint by $\vec\mu$.  The orphan enters the Gaussian action as the linear source $\vec\mu\cdot\vec C_0$: \emph{it couples to the liquid through the constrainer itself}.  This is the microscopic origin of the point source imposed by hand in Eq.~\eqref{eq:supp-strict-source}, and $\vec\mu$ plays exactly the role of the Gauss-law multiplier $\eta_I$ of Eq.~\eqref{eq:supp-lambda-solution}.

\paragraph*{Saddle point, renormalized moment and texture.}
The action~\eqref{eq:supp-hybrid-action} is quadratic in the spins and linear in $\vec\mu$, so its stationary point gives the exact mean of the liquid at fixed orphan direction $\vec n$; $\kappa$ keeps its undiluted value, the defect being a single site, and depends on $h$ only at order $h^2$.  Stationarity with respect to the spins, $\partial\mathcal S/\partial\vec S^a(\bq)^*=0$, has two sources, the uniform field at $\bq=\bm0$ and the multiplier through the constrainer,
\begin{equation}
\label{eq:supp-hybrid-spin-saddle}
 \sum_bh_{ab}(\bq)\,\langle\vec S^b(\bq)\rangle
 =\sqrt{N_{\rm uc}}\,\delta_{\bq,\bm0}\,\beta\vec h
 +\frac{e^{-\ii\bq\cdot\bR_0}}{\sqrt{N_{\rm uc}}}\,T^a_{l_0}(\bq)\,\vec\mu ,
\end{equation}
whose solution is, in real space,
\begin{equation}
\label{eq:supp-hybrid-mean}
 \langle\vec S^a_{\bR}\rangle=\langle\vec S^a\rangle_{{\rm ud},h}+\mathcal D^a(\bR)\,\vec\mu,
 \qquad
 \mathcal D^a(\bR)=\frac1{N_{\rm uc}}\sum_{\bq}
 e^{\ii\bq\cdot(\bR+\bm\tau_a-\bR_0)}\,
 \big[h^{-1}(\bq)\textbf T_{l_0}(\bq)\big]_a .
\end{equation}
The first term is the undiluted liquid in the field, uniform on each sublattice, $\langle\vec S^a\rangle_{{\rm ud},h}=\sum_bh^{-1}_{ab}(\bm0)\,\beta\vec h$; it is what the reference runs of Sec.~S8 measure and what the subtraction $\delta\langle\vec S\rangle=\langle\vec S\rangle-\langle\vec S\rangle_{{\rm ud},h}$ removes.  The second is the response to the multiplier, through $\mathcal D$, the SCGA propagator from the orphan constrainer to the spin at $(\bR,a)$.  Applying the constrainer of the cluster to Eq.~\eqref{eq:supp-hybrid-mean} gives
\begin{equation}
\label{eq:supp-sigma0}
 \vec C_0\big[\langle\vec S\rangle\big]=\langle\vec C_0\rangle_{{\rm ud},h}+\sigma_0^2\,\vec\mu,
 \qquad
 \sigma_0^2\equiv\sum_iC^i_{l_0}(\bR_0)\,\mathcal D_i
 =\frac{1}{N_{\rm uc}}\sum_{\bq}\sum_{ab}T^a_{l_0}(\bq)^*\,h^{-1}_{ab}(\bq)\,T^b_{l_0}(\bq):
\end{equation}
the value the constrainer already takes in the field, plus its response to its own source, $\sigma_0^2$ being the propagator from the constrainer back to itself.  Stationarity with respect to $\vec\mu$ restores the vacancy constraint, $\vec C_0[\langle\vec S\rangle]=m_{\rm O}\vec n$, and therefore fixes the multiplier,
\begin{equation}
\label{eq:supp-hybrid-mu-saddle}
 \sigma_0^2\,\vec\mu=m_{\rm O}\vec n-\langle\vec C_0\rangle_{{\rm ud},h},
\end{equation}
the relation behind Eq.~\eqref{eq:texture-factorization-prl} of the main text.  The multiplier supplies the difference between the value that the vacancy cluster imposes on the constrainer, $m_{\rm O}\vec n$, and the value that the field-polarized undiluted liquid already gives it, in units of the constrainer's response to its own source.  In zero field it reduces to $\vec\mu=m_{\rm O}\vec n/\sigma_0^2$.

In the SCGA these propagators are the connected correlation functions of the undiluted liquid, $\mathcal D^a(\bR)=\langle S^{a,z}_{\bR}C^z_0\rangle_{\rm ud}$ and $\sigma_0^2=\langle(C^z_0)^2\rangle_{\rm ud}$, the spin--charge and charge--charge correlators of the main text (the correlators $D_{12}$ and $D_1$ of Ref.~\cite{FloresCalderonBentonMoessner24}).  Equations~\eqref{eq:supp-hybrid-mean}--\eqref{eq:supp-hybrid-mu-saddle} are then the Gaussian conditional mean: in the undiluted liquid in the field $\vec C_0$ is Gaussian with mean $\langle\vec C_0\rangle_{{\rm ud},h}$ and variance $\sigma_0^2$ per component, and conditioning on $\vec C_0=m_{\rm O}\vec n$ shifts the mean of every spin by ${\rm Cov}(S^a_{\bR},C_0)/{\rm Var}(C_0)=\mathcal D^a(\bR)/\sigma_0^2$ times $m_{\rm O}\vec n-\langle\vec C_0\rangle_{{\rm ud},h}$.

The same Gaussian structure gives the distribution of the orphan direction.  Integrating out the spins and $\vec\mu$ in Eq.~\eqref{eq:supp-hybrid-partition} evaluates the Gaussian distribution of $\vec C_0$ at $m_{\rm O}\vec n$,
\begin{equation}
\label{eq:supp-orphan-weight}
 P(\vec n)\propto
 \exp\!\left[-\frac{\big(m_{\rm O}\vec n-\langle\vec C_0\rangle_{{\rm ud},h}\big)^2}{2\sigma_0^2}+\beta\vec h_{\rm O}\cdot\vec n\right]
 \propto\exp\!\left[\beta\big(\alpha\vec h+\vec h_{\rm O}\big)\cdot\vec n\right],
 \qquad
 \alpha=\frac{m_{\rm O}T\,\langle C^z_0\rangle_{{\rm ud},h}}{h\,\sigma_0^2},
\end{equation}
where we used $\vec n^{\,2}=1$: the term quadratic in $m_{\rm O}\vec n$ is constant on the unit sphere and exerts no torque, and only the cross term with the constrainer in the field orients the orphan.  Without the uniform field, $\langle\vec C_0\rangle_{{\rm ud},h}=0$ and the orphan is a free spin in $\vec h_{\rm O}$; this is the orphan-field protocol of Sec.~S8, with $\langle n^z\rangle=L(\beta h_{\rm O})$ exactly.  In the uniform field, $\vec h_{\rm O}=0$, the liquid itself exerts the effective field $\alpha\vec h$ on the orphan, and
\begin{equation}
\label{eq:supp-orphan-magnetization}
 \langle n^z\rangle=L\!\left(\frac{\alpha h}{T}\right),
 \qquad
 L(x)=\coth x-1/x,
\end{equation}
that of a free spin carrying the renormalized moment $\alpha$, the total moment of the orphan and its cloud per unit $\vec n$.  At low temperature every constrainer takes the same value in the field, $\langle C^z\rangle_{{\rm ud},h}=h/\bar cJ$, with $\bar c=\sum_lc_l$ the total constrainer weight of one site (minimize $\frac J2C^2-hC/\bar c$ per constrainer): $h/2J$ for kagome and checkerboard, where every spin belongs to two simplices, and $h/3(1+\gamma)J$ for the honeycomb-snowflake model.  With $\sigma_0^2\to T$, Eq.~\eqref{eq:supp-orphan-weight} gives $\alpha\to m_{\rm O}/\bar c$: $\alpha=1/2$ for kagome and checkerboard and $\alpha=m_{\rm O}/[3(1+\gamma)]$ for the honeycomb-snowflake model, equal to $2/9$ for a hexagon orphan at $\gamma=1/2$ and irrational whenever $\gamma$ is.  This is the ground-state result of Ref.~\cite{FloresCalderonBentonMoessner24}: in the ground-state manifold of the infinite lattice every constrainer except $\vec C_0$ vanishes, so $\sum_{\bR,l}\vec C_l(\bR)=\bar c\,\vec M_{\rm tot}=m_{\rm O}\vec n$.  At the simulated temperatures $\sigma_0^2$ lies slightly below $T$ and $\alpha$ slightly above $m_{\rm O}/\bar c$ (Sec.~S8).

Averaging Eq.~\eqref{eq:supp-hybrid-mean} over $\vec n$ with the weight~\eqref{eq:supp-orphan-weight} and subtracting the undiluted background gives $\delta\langle\vec S^a_{\bR}\rangle=\mathcal D^a(\bR)\langle\vec\mu\rangle$, which with Eq.~\eqref{eq:supp-hybrid-mu-saddle} is the general form of Eq.~\eqref{eq:texture-factorization-prl} of the main text.  By construction $\langle\vec C_0\rangle=m_{\rm O}\langle\vec n\rangle$: the hard constraint holds on average, and this fixes the amplitude.  The field enters only through two scalars, the orphan magnetization $\langle n^z\rangle$ and the $O(h)$ shift $\langle C^z_0\rangle_{{\rm ud},h}$; the spatial structure $\mathcal D^a(\bR)$ is that of the undiluted liquid in zero field, since $h_{ab}$ is independent of $h$ to linear order.  The contraction $h^{-1}\textbf T_{l_0}$ collapses by the push-through (Woodbury) identity,
\begin{equation}
\label{eq:supp-woodbury}
 h^{-1}(\bq)\,\mathsf T^{T}(\bq)
 =\mathsf T^{T}(\bq)\left[\kappa\,\mathbb{I}_{N_c}+\beta\,G(\bq)\right]^{-1},
 \qquad G_{ll'}=\textbf T_l^\dagger\textbf T_{l'},
\end{equation}
with $(\mathsf T^T)_{al}=T^a_l$ and $G$ the $N_c\times N_c$ Gram matrix of the constrainers, whose eigenvalues are the dispersive band energies.  For a single constrainer this is simply
\begin{equation}
\label{eq:supp-charge-spin-kernel}
 h^{-1}(\bq)\,\textbf T_{l_0}(\bq)
 =\frac{\textbf T_{l_0}(\bq)}{\kappa+\beta\,\omega(\bq)},
 \qquad \omega(\bq)=|\textbf T_{l_0}(\bq)|^2 ,
\end{equation}
which is the kernel of Eq.~\eqref{eq:texture-factorization-prl} of the main text and the explicit inverse used in Ref.~\cite{FloresCalderonBentonMoessner24}.  For $N_c>1$, Eq.~\eqref{eq:supp-woodbury} expands $h^{-1}\textbf T_{l_0}$ over the Gram eigenvectors; only the band that touches the flat band at $\bq_0$ is soft, and the gapped bands contribute terms that are analytic near $\bq_0$ and decay exponentially in real space.  At long distances Eq.~\eqref{eq:supp-charge-spin-kernel} therefore holds with $\textbf T_{l_0}$ replaced by its projection on the soft band, $U^*_{l_0s}$ times the normalized soft combination.

By Eq.~\eqref{eq:supp-woodbury}, the summand of Eq.~\eqref{eq:supp-sigma0} involves only the Gram eigenvalues $g_b(\bq)$, i.e.\ the dispersive band energies; the flat bands never enter, because $\textbf T_{l_0}$ has no weight on them.  With $G=U\,{\rm diag}(g_b)\,U^\dagger$,
\begin{equation}
\label{eq:supp-sigma0-bands}
 \sigma_0^2=\frac1{N_{\rm uc}}\sum_{\bq}\sum_b|U_{l_0b}(\bq)|^2\,\frac{g_b(\bq)}{\kappa+\beta g_b(\bq)},
 \qquad
 \frac{g}{\kappa+\beta g}=T-\frac{T\kappa}{\kappa+\beta g},
\end{equation}
where for a single constrainer $g=\omega(\bq)$ and $|U|^2=1$.  The second relation is an identity at every $\bq$.  Because $\sum_b|U_{l_0b}|^2=1$, its first term sums to exactly $T$, the leading equipartition value for the constrainer energy $\frac12(C^z_0)^2$, and the second term is the deficit by which the connected variance $\sigma_0^2=\langle(C^z_0)^2\rangle_{\rm ud}$ falls short of it, from momenta at which the constrainer is not stiff on the scale of $T$.  Away from $\bq_0$, where $\beta g_b\gg\kappa$, it is $\simeq\kappa T/g_b$, an $O(T)$ contribution of the whole zone that includes any gapped bands.  Near each of the $N_0$ touching points $\omega\simeq\lambda k^{2n}$, and the deficit is of order one for $|\bk|\lesssim\xi_n^{-1}$, weighted by the overlap $v(\bq_0) \equiv|U_{l_0s}(\bq_0)|^2$ of the orphan constrainer with the soft band $s$: $v(\bq_0)=1$ for a single constrainer and $v(\bq_0)=1/2$ on kagome, where the soft combination is $(M_{\mathcal A}-M_{\mathcal B})/\sqrt2$ (Table~\ref{tab:models}).  With $N_{\rm uc}^{-1}\sum_{\bq}\to A_{\rm uc}\int\dd^2k/(2\pi)^2$ and $u=(k\xi_n)^2$,
\begin{equation}
\label{eq:supp-sigma0-value}
 1-\frac{\sigma_0^2}{T}
 =\frac{v(\bq_0) N_0A_{\rm uc}}{4\pi\xi_n^2}\int_0^{(\Lambda\xi_n)^2}\frac{\dd u}{1+u^n}+O(T)
 =\begin{cases}
 \dfrac{v(\bq_0) N_0A_{\rm uc}}{4n\sin(\pi/n)\,\xi_n^2}+O(T), & n\ge2,\\[3mm]
 \dfrac{v(\bq_0) N_0A_{\rm uc}}{4\pi\xi_1^2}\,\ln(\Lambda\xi_1)^2+O(T), & n=1,
 \end{cases}
\end{equation}
with $\Lambda$ the momentum at which the continuum form of $\omega$ fails.  For $n\ge2$ the soft-region integral converges and the deficit is $O(\xi_n^{-2})\propto(\kappa T)^{1/n}$; at rank 1 it diverges logarithmically and the deficit is $O(T\ln T^{-1})$.  For the honeycomb-snowflake model ($n=2$, $N_0=2$ valleys, $v(\bq_0)=1$, $A_{\rm uc}=\sqrt3/2$) the first line gives $1-\sigma_0^2/T=0.0050$, $0.0087$ and $0.0159$ at $T/J=10^{-4}$, $3\times10^{-4}$ and $10^{-3}$, against $0.0050$, $0.0090$ and $0.0166$ from the full lattice sum; the difference is the $O(\kappa T)$ remainder.  For kagome the full sum gives $1-\sigma_0^2/T=0.073$, $0.040$ and $0.022$ at $T=0.1$, $0.05$ and $0.025$, falling slightly more slowly than $T$, as the logarithm implies.

\paragraph*{Continuum limit: the screened monopole.}
Near the band touching, project onto the soft modes of the charged sector, Eq.~\eqref{eq:soft-modes-prl} of the main text, $\vec\Phi_A(\bk)=\sum_a\mathcal U_{Aa}\vec S^a(\bq_0+\bk)$ with orthonormal rows, chosen so that $T^a_{l_0}(\bq_0+\bk)^*=d_A(\bk)\,\mathcal U_{Aa}+O(k^{n+1})$; then $\omega(\bq_0+\bk)=|\textbf d(\bk)|^2$, equal to $\lambda k^{2n}$ for the isotropic weight assumed throughout (Sec.~S4), and $\sum_a\mathcal U_{Aa}T^a_{l_0}=d_A^{\,*}$.  Converting $N_{\rm uc}^{-1}\sum_{\bq}\to A_{\rm uc}\int\dd^2k/(2\pi)^2$, with $A_{\rm uc}$ the unit-cell area, and normalizing the continuum soft modes as densities (spin per cell divided by $A_{\rm uc}$), the Fourier transform of Eq.~\eqref{eq:texture-factorization-prl} of the main text with the orphan at the origin is
\begin{equation}
\label{eq:supp-hybrid-continuum}
 \delta\langle\Phi^z_A(\bk)\rangle
 =\langle\mu^z\rangle\,\frac{d_A(\bk)^*}{\kappa+\beta|\textbf d(\bk)|^2}
 =Q_{\rm eff}\,\frac{d_A(\bk)^*}{\kappa T+|\textbf d(\bk)|^2},
 \qquad
 Q_{\rm eff}=T\langle\mu^z\rangle=\frac{T}{\sigma_0^2}\big(m_{\rm O}\langle n^z\rangle-\langle C^z_0\rangle_{{\rm ud},h}\big) ,
\end{equation}
which is the screened monopole of the main text; for $\bq_0\neq0$ a constant phase $e^{-\ii\bq_0\cdot\bR_0}$ has been absorbed.  Equation~\eqref{eq:supp-hybrid-continuum} thus rests on the SCGA and on steps (b) and (c) above; it is not exact for the diluted Heisenberg model, and Sec.~S8 compares the data with the lattice solution of step (a).  We distinguish the charge imposed on the orphan constrainer, $Q_{\rm core}\equiv m_{\rm O}\langle n^z\rangle-\langle C^z_0\rangle_{{\rm ud},h}$, from the monopole amplitude, $Q_{\rm eff}=T\langle\mu^z\rangle$ in the one-source approximation, with energies in units of $J$ ($T$ stands for $T/J$); there $Q_{\rm eff}=(T/\sigma_0^2)\,Q_{\rm core}$, which exceeds $Q_{\rm core}$ by $2$--$8\%$ at the rank-1 and by $0.5$--$1.7\%$ at the rank-2 temperatures of Sec.~S8.  Contracting with $d_A$ gives the induced charge density $Q_{\rm eff}|\textbf d|^2/(\kappa T+|\textbf d|^2)$: a point charge $Q_{\rm eff}$ at short distances and a compensating screening cloud of radius $\xi_n$.  For a harmonic symbol the same response is that of the gauge field, $\delta\langle\textbf B^z\rangle=L\,\delta\langle\bm\Phi^z\rangle$, and this charge is the source of its Gauss law, $\ii^{\,n}\textbf d_0\cdot\delta\langle\textbf B^z\rangle=\textbf d\cdot\delta\langle\bm\Phi^z\rangle$ (Sec.~S3).  The identification $Q_{\rm eff}\simeq Q_{\rm core}$ can also be read off the lattice without any continuum limit: on the infinite lattice, as $T\to0$, every constrainer other than $\vec C_0$ takes its undiluted value $\langle C^z\rangle_{{\rm ud},h}$, so after the subtraction of the undiluted background the lattice divergence through any loop around the orphan, within the screening length, is exactly $m_{\rm O}\vec n-\langle\vec C_0\rangle_{{\rm ud},h}$.  For $N_c>1$ the same holds with $d_A$ the symbol of the normalized soft combination of constrainers, to which the orphan constrainer couples with the overlap $U^*_{l_0s}$; in units of the staggered vertex charge of Sec.~S7 this overlap is compensated by the normalization of the soft combination, and the point charge is again $Q_{\rm eff}$ (Table~\ref{tab:models}).

 The stiffness is the curvature of the dispersive band, $\lambda=\lim_{k\to0}\omega(\bq_0+\bk)/k^{2n}$, and the screening length $\xi_n=(\lambda/\kappa T)^{1/2n}$ of the main text follows from the denominator alone.
The gauge charge is the orphan's own constrainer value: it saturates at $Q_{\rm core}\to m_{\rm O}-\langle C^z_0\rangle_{{\rm ud},h}$, independent of temperature, once the orphan is polarized, and in linear response, $L(x)\simeq x/3$, it is Curie-like,
\begin{equation}
\label{eq:supp-linear-response-charge}
 Q_{\rm core}\simeq\frac{m_{\rm O}\,\alpha\,h}{3T}\left(1-\frac{3T}{m_{\rm O}^2J}\right),
\end{equation}
the correction being the undiluted constrainer in the field, $\langle C^z_0\rangle_{{\rm ud},h}=h/\bar cJ$, relative to $m_{\rm O}\langle n^z\rangle$.
The gauge charge is thus not a free parameter but a measurable property of the vacancy cluster.

On a finite periodic lattice the total gauge charge vanishes identically: at rank 1 every link enters one vertex of each parity, so $\sum_vQ_v=0$, and at rank 2 the valley component of $\sum_{\hexagon}\vec C_{\hexagon}$ vanishes because the constrainer vanishes at $\pm\bm K$.  The response respects this, its induced charge density $Q_{\rm eff}|\textbf d|^2/(\kappa T+|\textbf d|^2)$ vanishing at $\bk=0$: the charge of the orphan is compensated by the screening cloud, which the finite-temperature response contains.  The strict point-charge solutions of Secs.~S4--S6 are infinite-plane solutions, valid at intermediate distances $a\ll r\ll\xi_n$, and the statements above that every other constrainer keeps its undiluted value hold only there.  The unscreened limit depends on the order of limits: for $L\to\infty$ at fixed $T$ the texture is the screened one, whereas for $T\to0$ at fixed $L$ the compensating charge spreads over the whole torus once $\xi_n\gtrsim L$, which matters most for the nondecaying rank-2 and the growing $n\ge3$ fields.  The simulations of Sec.~S8 have $L\ge5\,\xi_n$.

\subsection*{S3. Topological invariants and the gauge field}\phantomsection\label{sec:S3}

\paragraph*{Why two dimensions, and complex notation.}
The harmonic polynomials of degree $n$ in two variables span a two-dimensional space for every $n\ge1$, with basis $\operatorname{Re}(k_x+\ii k_y)^n$ and $\operatorname{Im}(k_x+\ii k_y)^n$.  The charged sector of a symmetric traceless rank-$n$ field in $d=2$ is therefore two-component at every rank, its normalized symbol is a map $S^1\to S^1$, and the invariant lives in $\pi_1(S^1)=\mathbb Z$; a general constrainer can have a larger charged sector (Sec.~S1), which (H1) of Sec.~S4 excludes.  This is what makes a rank-$n$ statement possible in $d=2$ .

We write
\begin{equation}
\label{eq:supp-complex-notation}
 \zeta=k_x+\ii k_y=ke^{\ii\phi},
 \quad
 u=e^{\ii\phi},
 \quad
 z=x+\ii y=re^{\ii\theta},
\end{equation}
and combine the two components of any real two-component object into a single complex one, $d_c=d_1+\ii d_2$, $\Phi_c=\Phi_1+\ii\Phi_2$ and $B_c=B_1+\ii B_2$.  The canonical rank-$n$ symbol is then simply $d_{0,c}(\bk)=\zeta^n$, and the canonical rank-$n$ Gauss law is $\operatorname{Re}\big[\bar\partial_c^{\,n}B_c\big]=\rho$ with $\bar\partial_c=\partial_x-\ii\partial_y$.  Degrees are computed as winding numbers of complex-valued maps, $\deg f=\frac{1}{2\pi}\oint\dd\arg f$.

In this notation both invariants of the main text are degrees of maps $S^1\to S^1$.  The bulk invariant, Eq.~\eqref{eq:general-framework-prl}, is $W_{\rm bulk}=\deg\hat d_c$ on a loop around $\bq_0$, and the defect invariant, Eq.~\eqref{eq:defect-winding-prl}, is $W_{\rm defect}=\deg \Phi_c$ on a circle around the orphan.  Both live in $\pi_1(S^1)=\mathbb Z$ once the real structure of Sec.~S1 has made the charged sector two-component and real.

\paragraph*{Gauge field and soft modes.}
The full contraction of a symmetric traceless rank-$n$ tensor with a vector is
\begin{equation}
\label{eq:supp-contraction}
 k_{\mu_1}\cdots k_{\mu_n}B_{\mu_1\cdots\mu_n}
 =\textbf d_0(\bk)\cdot\textbf B,
 \qquad
 \textbf d_0(\bk)=\big(\operatorname{Re}\zeta^n,\operatorname{Im}\zeta^n\big)=k^n(\cos n\phi,\sin n\phi).
\end{equation}
Both sides are homogeneous polynomials of degree $n$ in $\bk$, harmonic because $B$ is traceless, and such a polynomial is fixed by its value and its $k_y$ derivative at $\bk=\hat{\bm x}$, which are $B_1$ and $nB_2$ on both sides.  For real components the contraction equals $\operatorname{Re}[\bar\zeta^{\,n}B_c]$, the form used in real space below; for complex Fourier amplitudes only $\textbf d_0\cdot\textbf B$ applies.  With $\partial_\mu\leftrightarrow\ii k_\mu$ the rank-$n$ Gauss law, Eq.~\eqref{eq:rankn-gauss-prl} of the main text with a source, therefore reads
\begin{equation}
\label{eq:supp-canonical-gauss}
 \partial_{\mu_1}\cdots\partial_{\mu_n}B_{\mu_1\cdots\mu_n}=\operatorname{Re}\big[\bar\partial_c^{\,n}B_c\big]=\rho
 \quad\Longleftrightarrow\quad
 \ii^{\,n}\,\textbf d_0(\bk)\cdot\textbf B(\bk)=\rho(\bk),
\end{equation}
with the canonical symbol $\textbf d_0$, of degree $n$.  The microscopic constraint, on the other hand, is $\textbf d(\bk)\cdot\bm\Phi(\bk)=\rho(\bk)$ in terms of the soft modes (Sec.~S2).  The two are the same equation if and only if
\begin{equation}
\label{eq:supp-L-def}
 B_A=L_{AB}\Phi_B,
 \qquad
 \textbf d(\bk)=\ii^{\,n}L^T\textbf d_0(\bk),
\end{equation}
with a constant $2\times2$ matrix $L$, which is real because $\ii^{-n}\textbf d$ is (Sec.~S4).  Since the components of $\textbf d_0$ form a basis of the harmonic polynomials of degree $n$, $L$ exists, and is unique, exactly when the two components of $\textbf d$ are harmonic; it is then invertible exactly when $\textbf d$ is elliptic.  The soft modes are the gauge field in a rotated and rescaled basis, and the data of the theory transform covariantly:
\begin{itemize}[leftmargin=1.6em]
 \item \emph{Windings.}  $\ii^{-n}\textbf d=L^T\textbf d_0$ gives $W_{\rm bulk}=n\,\sgn\det L$: in the polar decomposition $L=OP$ the positive matrix $P$ is deformed to the identity without closing the zero, while the orthogonal $O$ preserves or reverses the orientation according to $\det O=\sgn\det L$.  Likewise a texture $\bm\Phi=L^{-1}\textbf B$ winds $\sgn\det L$ times as often as $\textbf B$.
 \item \emph{Metric.}  $\frac12\bm\Phi\cdot\mathcal K\bm\Phi=\frac12\textbf B\cdot\mathcal K_B\textbf B$ with $\mathcal K_B=L^{-T}\mathcal KL^{-1}$, positive definite exactly when $\mathcal K$ is.
 \item \emph{Stiffness.}  $|\textbf d|^2=\textbf d_0\cdot LL^T\textbf d_0$ equals $\lambda k^{2n}$ for every $\bk$ exactly when $LL^T=\lambda\mathbb I$, i.e.\ $L=\sqrt\lambda\,O$ with $O$ orthogonal.  The SCGA weight of the soft modes, $\frac12\int_{\bk}\big[\kappa T|\bm\Phi|^2+|\textbf d\cdot\bm\Phi|^2\big]$ (Sec.~S2), then becomes $\frac1{2\lambda}\int_{\bk}\big[\kappa T|\textbf B|^2+\lambda|\textbf d_0\cdot\textbf B|^2\big]$.
\end{itemize}
Every rank-1 symbol is harmonic: $\textbf d=\ii\mathcal M\bk$ gives $L=\mathcal M^T$ and the flux $B_\mu=\mathcal M_{A\mu}\Phi_A$ of Sec.~S5, with $LL^T=\mathcal M^T\mathcal M=\lambda\mathbb I$ on kagome and checkerboard.  The honeycomb-snowflake symbol is $\textbf d=\sqrt\lambda\,\textbf d_0$ with $\lambda=9/32$, so that $L=-\sqrt\lambda\,\mathbb I$, the sign compensating $\ii^2$: $\textbf B=-\sqrt\lambda\,\bm\Phi$, $\det L>0$ and $W_{\rm bulk}=2$ (Sec.~S7).  At rank $n\ge2$, by contrast, a symbol can be elliptic, and even of maximal degree, without being harmonic, when it contains the modes $|m|\le n-2$ that carry a factor $k^2=\zeta\bar\zeta$; then no constant $L$ exists, the constraint is not the Gauss law of a symmetric traceless tensor, and the stiffness $|\textbf d|^2$ is anisotropic.  The isotropic Gaussian weight assumed in Sec.~S4 excludes such symbols, a symbol of maximal degree then being harmonic with $LL^T=\lambda\mathbb I$; they are discussed with the anisotropic case at the end of that section.

\paragraph*{Orientation.}
Both invariants are degrees of maps into the same charged plane, so an orientation-reversing change of its basis, such as $\Phi_2\to-\Phi_2$ or the exchange of the valleys $\pm\bq_0$, reverses both: the equality $W_{\rm defect}=W_{\rm bulk}$ is basis independent, while the sign of each is a convention and $|W|=n$ is the intrinsic invariant.  The gauge field provides the convention used here.  Its components $B_1=B_{xx\cdots x}$ and $B_2=B_{xx\cdots xy}$ refer to a right-handed spatial frame, in which the canonical symbol $\textbf d_0$ has degree $+n$, and the map from the spins to $\textbf B$ is fixed by the constrainer, as the oriented link flux at rank 1; $W_{\rm bulk}=n\,\sgn\det L$ then records the orientation of the soft-mode basis relative to this oriented spin-to-flux map, and the bases of Sec.~S7 have $\det L>0$ on all three lattices.  In the language of Refs.~\cite{BentonMoessner21,YanClassification}, $W$ is the local index, or vorticity, of an isolated zero of the constrainer: a property of a single band touching, not a global invariant of the constrainer bands, and distinct from their fragile or other band topology.

\subsection*{S4. The rank-$n$ bulk--defect correspondence: proof and conditions}\phantomsection\label{sec:S4}

The rank-1 and rank-2 textures of Secs.~S5 and S6 share a single mechanism, which we prove here at general rank.  The essential point is that in two dimensions the Fourier transform acts \emph{diagonally} on angular harmonics.  It can therefore reweight the angular channels of a field but can never mix them, so an integer degree carried by a momentum-space symbol is transferred to real space unless a subleading channel is promoted by an infrared divergence.  Hypotheses (H2) and (H3) below are sufficient to forbid this.

\paragraph*{Conditions.}
Fix $n\ge1$.  The theorem holds under the four conditions stated in the main text, which are sufficient rather than necessary; the isotropy in (H3) is relaxed at the end of this section:
\begin{enumerate}[label=(H\arabic*),leftmargin=2.4em]
 \item \emph{Two-component real structure and ellipticity.}  The charged sector, the span of the vectors $\textbf d(\bk)$, is two-dimensional, which is automatic at $n=1$ and a hypothesis for $n\ge2$ (Sec.~S1).  In the real basis fixed by $\Theta$, $\ii^{-n}\textbf d(\bk)$ is a real homogeneous polynomial of degree $n$, i.e.\ a symbol of derivative order $n$, the factor $\ii^{\,n}$ being that of $\partial_\mu\leftrightarrow\ii k_\mu$, and $\textbf d(\bk)\ne0$ for $\bk\ne0$.  Without two components and the real structure the charged sector is not a real plane and the circle-map degree is not defined; without ellipticity the symbol vanishes on the loop, the touching is a line rather than a point, and no degree is defined.  A constant phase rotates the texture rigidly, so in the rest of this section $\textbf d$ denotes the real, phase-stripped symbol $\ii^{-n}\textbf d$ of Secs.~S2 and S3, and we write $\textbf d(\bk)=k^n\textbf a(\phi)$ with the angular amplitude $\textbf a(\phi)=\ii^{-n}\textbf d(\bk)/k^n$ (in terms of the symbol of Secs.~S2 and S3) nowhere zero; the real unit vector of the main text is $\hat{\textbf d}=\textbf a/|\textbf a|$, of the same winding, and $a_c=a_1+\ii a_2$; the phase $\ii^{\,n}$ is restored in the gauge-field form below.
 \item \emph{Maximal vorticity,} $|W_{\rm bulk}|=|\deg\hat{\textbf d}|=n$.  In the notation of Sec.~S3, $d_c=\sum_{j=0}^np_j\zeta^j\bar\zeta^{\,n-j}$, so that $a_c(u)=u^{-n}P(u^2)$ with $P(w)=\sum_jp_jw^j$; the terms $j=n$ and $j=0$ are the harmonic modes $m=\pm n$, and every other term carries a factor $\zeta\bar\zeta=k^2$.  By the argument principle $\deg\hat{\textbf d}=2Z(P)-n$, with $Z(P)$ the number of zeros of $P$ in $|w|<1$, so $\deg\hat{\textbf d}=+n$ exactly when $P$ has degree $n$ and all its zeros in the unit disc; $-n$ is the mirror statement.  When the modes $|m|\le n-2$ lower the degree they populate the infrared channels of the Fourier transform (Lemma below), which can then dominate the texture and decouple its winding from $\deg\hat{\textbf d}$.  Maximal vorticity is sufficient, not necessary: $\textbf d=k^2(k_x,k_y)$ has derivative order 3 and winding 1, is isotropic, and its screened response, a single harmonic $m=1$, winds once wherever its radial amplitude is nonzero.
 \item \emph{Positive isotropic Gaussian weight.}  The quadratic functional on the charged sector, $\mathcal F_n=\frac12\int\dd^2r\,\Phi_{AI}\mathcal K_{AB}\Xi_{IJ}\Phi_{BJ}$, has the spectator metric $\Xi$ symmetric positive definite and $\mathcal K=\kappa\mathbb I$, and the pinch point is isotropic, $|\textbf d(\bk)|^2=\lambda k^{2n}$, so that the Gaussian weight of the soft modes, $\frac12\int_{\bk}\big[\kappa T|\bm\Phi|^2+|\textbf d\cdot\bm\Phi|^2\big]$ (Sec.~S3), is rotation invariant.  Positivity makes the constrained saddle unique; the factorization over spectator labels makes every spin component or valley see the same spatial metric, so that $\Xi$ cancels from the texture (Sec.~S5).  The SCGA has $\mathcal K=\kappa\mathbb I$ and $\Xi=\mathbb I$, and lattice symmetry makes its pinch point isotropic for the three models studied here (Corollary below).
 \item \emph{Monopolar orphan and invertible reconstruction.}  The defect carries a nonzero charge in the same Gauss-law sector, $\rho=Q\delta^{(2)}(\br)$ with $Q\ne0$, and the map from the measured spins to the charged field is invertible: the projection $\mathcal U$ of Eq.~\eqref{eq:soft-modes-prl} on the charged sector and $\textbf B=L\bm\Phi$ with $\det L\ne0$.  If the monopole cancels, the leading multipole sources angular channels of degree $n\pm1$, which is why the charge is measured independently of the winding (Sec.~S8); and when several spectator sectors are charged, the winding is that of a fixed contraction $v_I\Phi_{AI}$ with $v_IQ_I\ne0$.
\end{enumerate}
Under (H3), (H2) fixes the symbol completely: $|a_c|$ is constant on $|u|=1$ only if $P$ is a monomial, $P\propto w^j$, of degree $2j-n$, so (H2) requires $j=n$ or $j=0$,
\begin{equation}
\label{eq:supp-isotropic-symbol}
 a_c(u)=\sqrt\lambda\,e^{\ii\chi}u^{\pm n},
\end{equation}
with a constant phase $\chi$.  The symbol is harmonic, with $LL^T=\lambda\mathbb I$ in Eq.~\eqref{eq:supp-L-def}.  Minimizing $\mathcal F_n$ subject to the Gauss law gives the same Lagrange-multiplier saddle as Eq.~\eqref{eq:supp-strict-fourier-solution}, now with the rank-$n$ symbol,
\begin{equation}
\label{eq:supp-rankn-saddle}
 \hat{\bm\Phi}(\bk)=-Q\,\frac{\textbf d(\bk)}{|\textbf d(\bk)|^2}=k^{-n}\bm\varepsilon(\phi),
 \qquad
 \varepsilon_c=-\frac{Q}{\overline{a_c}}=-\frac{Q\,e^{\ii\chi}}{\sqrt\lambda}\,u^{\pm n},
\end{equation}
a single angular harmonic.

\textbf{Lemma (angular transfer).}\quad If $\hat\Phi_c(\bk)=k^{-n}\varepsilon(\phi)$, with Fourier coefficients $\varepsilon_m$, then, for $r>0$,
\begin{equation}
\label{eq:supp-hecke-bochner}
 \Phi_c(\br)=\frac{r^{\,n-2}}{2\pi}
 \sum_m \ii^{\,m}\,\mathcal I(m,n)\,\varepsilon_m\,e^{\ii m\theta},
 \qquad
 \mathcal I(m,n)=\int_0^\infty\!\dd u\,u^{1-n}J_m(u).
\end{equation}
\emph{Proof.}  Insert the Jacobi--Anger expansion $e^{\ii\bk\cdot\br}=\sum_l\ii^{\,l}J_l(kr)e^{\ii l(\phi-\theta)}$ into $\Phi_c=\int\!\frac{\dd^2k}{(2\pi)^2}e^{\ii\bk\cdot\br}\hat\Phi_c$.  The angular integral $\int_0^{2\pi}\!\dd\phi\,e^{\ii(l+m)\phi}=2\pi\delta_{l,-m}$ selects $l=-m$, leaving $\ii^{-m}J_{-m}(kr)e^{\ii m\theta}$; using $J_{-m}=(-1)^mJ_m$ and rescaling $u=kr$ gives Eq.~\eqref{eq:supp-hecke-bochner}. $\square$

The angular index $m$ is conserved: the transform is a \emph{multiplier} in mode space.  Evaluating the radial integral \cite{SteinWeiss} gives, for $|m|\ge n$ and $|m|\equiv n\ (\mathrm{mod}\ 2)$,
\begin{equation}
\label{eq:supp-radial-multiplier}
 \mathcal I(m,n)=\sigma_m\,2^{\,1-n}\,
 \frac{\Gamma\!\left(\frac{|m|-n+2}{2}\right)}
      {\Gamma\!\left(\frac{|m|+n}{2}\right)},
 \qquad
 \sigma_m=\begin{cases}1,&m\ge0,\\(-1)^m,&m<0,\end{cases}
\end{equation}
so that $\ii^{\,m}\mathcal I(m,n)=\ii^{\,n}\nu_s$ for $|m|=n+2s$, with $\nu_s=(-1)^s2^{1-n}s!/(s+n-1)!$ independent of the sign of $m$.  For $|m|\le n-2$ the integral \emph{diverges} at small $u$, because $J_m(u)\sim(u/2)^{|m|}/|m|!$ makes the integrand behave as $u^{1-n+|m|}$.  We call these the \emph{infrared channels}; they are logarithmic for $|m|=n-2$, cutoff-power divergent for $|m|<n-2$, and absent for $n=1$.

\textbf{Theorem (rank-$n$ bulk--defect correspondence).}\quad Under (H1)--(H4) the orphan texture is
\begin{equation}
\label{eq:supp-rankn-texture}
 \Phi_c(\br)=-\frac{Q\,\ii^{\,n}e^{\ii\chi}}{2^n\pi\,(n-1)!\,\sqrt\lambda}\,r^{\,n-2}e^{\pm\ii n\theta},
\end{equation}
and
\begin{equation}
\label{eq:supp-rankn-result}
 W_{\rm defect}=W_{\rm bulk}=\deg\hat{\textbf d}=\pm n .
\end{equation}
\emph{Proof.}  By Eq.~\eqref{eq:supp-rankn-saddle} $\varepsilon$ has the single mode $m=\pm n$, which is not an infrared channel, so the Lemma applies with $\ii^{\pm n}\mathcal I(\pm n,n)=\ii^{\,n}\nu_0$, $\nu_0=2^{1-n}/(n-1)!$, and yields Eq.~\eqref{eq:supp-rankn-texture}.  Its phase winds $\pm n$ times, which is $\deg\hat{\textbf d}=W_{\rm bulk}$, and $W_{\rm defect}$ is unchanged by the invertible reconstruction (H4), whose constant matrix multiplies degrees by $\sgn\det$ on both sides simultaneously (Sec.~S3). $\square$

For the honeycomb-snowflake model $a_c=-\sqrt\lambda\,u^2$ ($\chi=\pi$, Sec.~S3), and Eq.~\eqref{eq:supp-rankn-texture} is $\Phi_c=-\frac{Q}{4\pi\sqrt\lambda}e^{2\ii\theta}$, i.e.\ the canonical quadrupole $B_c=-\sqrt\lambda\,\Phi_c=\frac{Q}{4\pi}e^{2\ii\theta}$ of Eq.~\eqref{eq:supp-rank2-exact-saddle}.

By Eq.~\eqref{eq:supp-isotropic-symbol}, $L=\sqrt\lambda\,O$ with $O$ orthogonal, and $\textbf B=L\bm\Phi$ obeys the canonical Gauss law, Eq.~\eqref{eq:supp-canonical-gauss}, whose angular amplitude $d_{0,c}/k^n=u^n$ is a single angular mode, with the isotropic metric $\mathcal K_B=L^{-T}\mathcal KL^{-1}=(\kappa/\lambda)\mathbb I$.  The saddle, keeping the phase $\ii^{\,n}$, is $\hat{\textbf B}=Q\,\ii^{-n}\textbf d_0/k^{2n}$, and the Lemma gives the field of a rank-$n$ point charge,
\begin{equation}
\label{eq:supp-canonical-point-charge}
 B_c(\br)=\frac{Q\,r^{\,n-2}}{2^n\pi\,(n-1)!}\,e^{\ii n\theta},
\end{equation}
which winds $n$ times.  The soft modes $\bm\Phi=L^{-1}\textbf B$ then wind $n\,\sgn\det O$ times, which is $W_{\rm bulk}$ by Eq.~\eqref{eq:supp-L-def}: both invariants are the degree $n$ of the canonical Gauss law multiplied by the orientation of the constant map between soft modes and gauge field.

\paragraph*{Finite temperature and symmetry.}
Equation~\eqref{eq:texture-factorization-prl} of the main text and Eq.~\eqref{eq:supp-charge-spin-kernel} make the theorem applicable to the physical finite-temperature response and not only to the strict saddle.  The texture factorizes: its entire $\bR$ dependence is the spin--charge correlator of the undiluted liquid, which is the constrainer divided by the strictly positive scalar $\kappa+\beta\omega(\bq)$, and the orphan contributes only the scalar $\langle\mu^z\rangle=Q_{\rm core}/\sigma_0^2$ (one-source approximation of Sec.~S2), whose nonvanishing is (H4).  For an isotropic pinch point the screening factor $\lambda k^{2n}/(\kappa T+\lambda k^{2n})$ of Eq.~\eqref{eq:supp-hybrid-continuum} depends on $k$ only, so it multiplies the single harmonic of Eq.~\eqref{eq:supp-rankn-saddle} by a radial function, and the Lemma gives Eq.~\eqref{eq:supp-rankn-texture}, with $Q=Q_{\rm eff}$ and up to the constant phase $-\ii^{-n}$ of the physical response [Eq.~\eqref{eq:em-texture} of the End Matter], times the envelope
\begin{equation}
\label{eq:supp-rankn-envelope}
 F_n\!\left(\frac r{\xi_n}\right),\qquad
 F_n(x)=2^{n-1}(n-1)!\int_0^\infty\!\dd u\,\frac{u^{n+1}J_n(u)}{u^{2n}+x^{2n}},
\end{equation}
with $F_n\to1$ for $r\ll\xi_n$; $F_1(x)=xK_1(x)$ and $F_2$ are those of Secs.~S5 and S6.  Temperature therefore enters only through the envelope, never through the angular map, and $W_{\rm defect}=W_{\rm bulk}$ on every circle on which $F_n\ne0$, in particular throughout the window $a\ll R\ll\xi_n$.  For $n\ge2$ the strict solution, of envelope $r^{\,n-2}$, is an intermediate-distance form, and $\xi_n$ supplies its infrared cutoff.  The winding is a property of the undiluted spin liquid, measured with a defect.

\textbf{Corollary (symmetry).}\quad Under a lattice rotation by $2\pi/p$ the rank-$n$ charged doublet rotates by $2\pi n/p$.  A quadratic form invariant under a rotation by $2\pi n/p$ is isotropic unless $2\pi n/p\in\pi\mathbb Z$, and with a $C_p$ rotation $a_c$ retains only the modes $m\equiv n\ (\mathrm{mod}\ \mathrm{lcm}(2,p))$, so that $|a_c|^2$ retains only multiples of $\mathrm{lcm}(2,p)$ among $|m|\le2n$.  Hence a $C_p$ rotation with
\begin{equation}
\label{eq:supp-symmetry-protection}
 \mathrm{lcm}(2,p)>2n ,
\end{equation}
which implies $p\nmid2n$, forces the whole Gaussian weight to be isotropic, as (H3) requires.  This covers kagome ($n=1$, $C_6$), checkerboard ($n=1$, $C_4$) and the honeycomb-snowflake model ($n=2$, $C_3$).

\paragraph*{Anisotropic weights.}
Without (H3) the correspondence holds under weaker conditions.  A general positive metric is removed by $\bm\Phi'=\mathcal K^{1/2}\bm\Phi$, which turns the saddle $\hat{\bm\Phi}=-Q\,\mathcal K^{-1}\textbf d/(\textbf d\cdot\mathcal K^{-1}\textbf d)$ into Eq.~\eqref{eq:supp-rankn-saddle} with the symbol $\textbf d'=\mathcal K^{-1/2}\textbf d$; since $\det\mathcal K^{1/2}>0$, neither invariant changes, and only an anisotropic stiffness $|\textbf d'(\bk)|^2=k^{2n}|\textbf a'(\phi)|^2$ matters.  By the root count of (H2), degree $+n$ means $d'_c=p_n\prod_{i=1}^n(\zeta-z_i\bar\zeta)$ with all $|z_i|<1$.  (i)~The strict saddle obeys the correspondence for every such symbol.  All factors $\bar\zeta-\bar z_i\zeta=\bar\zeta(1-\bar z_iu^2)$ lie in one half plane, so Feynman parameters give $\hat\Phi'_c=-Q/\overline{d'_c}=-(Q/\bar p_n)(n-1)!\int_\Delta\dd\tau\,(\bar\zeta-\bar z_\tau\zeta)^{-n}$, with $z_\tau=\sum_i\tau_iz_i$ on the simplex $\Delta$; expanding in $u^2$, the binomial coefficients cancel the weights $\nu_s$, and the Lemma gives
\begin{equation}
\label{eq:supp-anisotropic-texture}
 \Phi'_c(\br)=-\frac{Q\,\ii^{\,n}r^{\,n-2}e^{\ii n\theta}}{2^n\pi\,\bar p_n}\int_\Delta\frac{\dd\tau}{1+\bar z_\tau e^{2\ii\theta}} .
\end{equation}
Since $|z_\tau|<1$ the integrand has real part above $\frac12$, so the average neither vanishes nor winds: $W_{\rm defect}=n$, and the phase stays within $\pi/2$ of $n\theta$; degree $-n$ follows by complex conjugation.  (ii)~At finite temperature the screening factor $\varsigma=|\textbf d'|^2/(\kappa T+|\textbf d'|^2)$ depends on $\phi$, mixes angular harmonics and populates the infrared channels $|m|\le n-2$.  Write $d'_c=k^na'_c(\phi)$.  Since $1/\overline{a'_c}$ contains only the modes $m=n+2s$, $s\ge0$, the channel $m$ of $\hat\Phi'_c=-Q\varsigma/(k^n\overline{a'_c})$ with $|m|<n$ comes from $\varsigma-1$ alone, which is concentrated at $k\lesssim\xi_n^{-1}$, and in the window it adds a term of relative order $(\xi_n/r)^{n-2-|m|}$ that does not vanish.  At $n=1$ there is no such channel, and in the sheared coordinates in which the stiffness is isotropic the screened monopole is that of Sec.~S5, radial with unit winding, for every anisotropy.  At $n=2$ the only channel is $m=0$.  With $\langle1/\overline{a'_c}\rangle_\phi=0$ and $\int_0^\infty\frac{\dd k}{k}\big[(1+bk^4)^{-1}-\Theta(1-k)\big]=-\frac14\ln b$, it contributes for $r\ll\xi_2$ the temperature-independent constant $-(Q/8\pi)\langle\ln|a'_c|^2/\overline{a'_c}\rangle_\phi$, and for $a'_c=p_2u^2(1-z_1\bar u^2)(1-z_2\bar u^2)$ the texture in the window is
\begin{equation}
\label{eq:supp-anisotropic-rank2}
 \Phi'_c=\frac{Q}{4\pi\bar p_2(\bar z_1-\bar z_2)}\,\ln\!\Big[\Lambda_0\,\frac{1+\bar z_1w}{1+\bar z_2w}\Big],
 \qquad
 \Lambda_0=\Big(\frac{1-|z_2|^2}{1-|z_1|^2}\Big)^{1/2}\frac{1-z_1\bar z_2}{|1-z_1\bar z_2|},
\end{equation}
with $w=e^{2\ii\theta}$; $\Lambda_0$ is the infrared constant, and without it Eq.~\eqref{eq:supp-anisotropic-rank2} is the $n=2$ case of Eq.~\eqref{eq:supp-anisotropic-texture}.  For $|z_i|<1$ the argument of the logarithm is analytic and nonzero in $|w|\le1$, so the branch fixed at $w=0$ is single valued there, and the winding of $\Phi'_c$ in $w$ is the number of solutions of $\Lambda_0(1+\bar z_1w)=1+\bar z_2w$ in the disc.  This linear equation has one solution, which lies inside because $|1-z_1\bar z_2|^2-|z_1-z_2|^2=(1-|z_1|^2)(1-|z_2|^2)>0$; hence $W_{\rm defect}=2$ for every anisotropy.  For coincident roots, $z_1=z_2=z$, Eq.~\eqref{eq:supp-anisotropic-rank2} tends to the Blaschke factor $\Phi'_c=Q\,(w+z)/[4\pi\bar p_2(1-|z|^2)(1+\bar zw)]$, of constant modulus on every circle and again of winding 2.  For $n\ge3$ the correspondence holds when a $C_p$ rotation with $\mathrm{lcm}(2,p)\ge2n$ forbids the infrared channels.  Otherwise they enter the window with a relative amplitude of order one at $n=3$, comparable to the harmonic $m=\pm3$, and of order $(\xi_n/r)^2$ or larger for $n\ge4$, so that for strong or generic anisotropy the winding is no longer fixed by $\deg\hat{\textbf d}$; we do not establish its value there. These statements concern general two-component symbols of derivative order $n$,
including nonharmonic symbols. Constant metric preserves harmonicity,
although it can make the stiffness anisotropic. We checked Eq.~\eqref{eq:supp-anisotropic-texture} against a direct evaluation of the Lemma for $n\le5$, and Eq.~\eqref{eq:supp-anisotropic-rank2} against a two-dimensional fast Fourier transform of the screened texture on a $4096^2$ grid ($\xi_2\simeq56$ and $100$), which it matches within $2\%$ at $R=8$ and $16$, against $13\%$ without $\Lambda_0$.

\subsection*{S5. Rank 1: Coulomb spin liquids}\phantomsection\label{sec:S5}

At rank 1 the symbol is linear, $d_A(\bk)=\ii\mathcal M_{A\mu}k_\mu+O(k^2)$ with $\det\mathcal M\ne0$, and the gauge field is the flux
\begin{equation}
\label{eq:supp-flux-reconstruction}
 B_{\mu I}=\mathcal M_{A\mu}\Phi_{AI},
\end{equation}
i.e.\ $L=\mathcal M^T$ in Eq.~\eqref{eq:supp-L-def}, which turns the projected constraint into $\ii k_\mu B_{\mu I}$, the Fourier transform of $\partial_\mu B_{\mu I}$, with $W_{\rm bulk}=\sgn\det\mathcal M$.  If $\det\mathcal M=0$ the spins do not determine a two-component flux and no winding equality follows.  For kagome and checkerboard $\mathcal M\propto\mathbb I$ (Sec.~S7).  We use
\begin{equation}
\label{eq:supp-fourier-convention}
 B_{\mu I}(\br)=\int_{\bq}e^{\ii\bq\cdot\br}B_{\mu I}(\bq),
 \qquad
 \int_{\bq}\equiv\int{\dd^2q\over(2\pi)^2},
 \qquad \partial_\mu\longleftrightarrow\ii q_\mu .
\end{equation}

\paragraph*{Metric and pinch point.}
At rank 1 the functional of (H3) is
\begin{equation}
\label{eq:supp-factorized-action}
 \mathcal F_0[B]={1\over2}\int\dd^2r\,
 B_{\mu I}K^{(s)}_{\mu\nu}\Xi_{IJ}B_{\nu J},
 \qquad K^{(s)}>0,\quad \Xi>0,
\end{equation}
where the spatial stiffness $K^{(s)}$ sets the ellipticity of the pinch point and of the defect envelope, and the spectator metric $\Xi$ the relative weights of spin components, valleys or copies.  Neither is the constrainer: the constrainer fixes the Gauss-law symbol and its degree, while $K^{(s)}$ and $\Xi$, the positive SCGA Hessian projected on the charged subspace, fix nonuniversal shapes and amplitudes.  Condition (H3) takes $K^{(s)}\propto\mathbb I$, as in the SCGA below; we keep $K^{(s)}$ general because at rank 1 the correspondence holds for every anisotropy (Sec.~S4).  With $\mathsf A\equiv K^{(s)-1}$ and $H\equiv\Xi^{-1}$, imposing $q_\mu B_{\mu I}=0$ with one multiplier per spectator sector and using fluctuation--response gives the transverse pinch point
\begin{equation}
\label{eq:supp-constrained-covariance}
 \langle B_{\mu I}(\bq)B_{\nu J}(-\bq)\rangle_{{\rm ud},c}
 =T H_{IJ}\left[\mathsf A_{\mu\nu}-{(\mathsf A\bq)_\mu(\mathsf A\bq)_\nu\over\bq\cdot\mathsf A\bq}\right].
\end{equation}
In the SCGA, in the normalization of Sec.~S3, $K^{(s)}=(\kappa T/\lambda)\mathbb I$ and $\Xi=\mathbb I$, and at finite temperature the longitudinal part is suppressed rather than projected out,
\begin{equation}
\label{eq:supp-finite-T-covariance}
 \langle B_\mu(\bq)B_\nu(-\bq)\rangle_{\rm ud}
 ={\lambda\over\kappa}P^{\rm T}_{\mu\nu}
 +{\lambda T\over\kappa T+\lambda q^2}P^{\rm L}_{\mu\nu},
 \qquad
 P^{\rm L}_{\mu\nu}=\frac{q_\mu q_\nu}{q^2},\quad P^{\rm T}=\mathbb I-P^{\rm L},
\end{equation}
so the pinch point is rounded below $q\sim\xi_1^{-1}$.

\paragraph*{Strict monopole.}
For a localized gauge charge,
\begin{equation}
\label{eq:supp-strict-source}
 \ii q_\mu B_{\mu I}(\bq)=Q_{{\rm eff},I},
\end{equation}
extremizing $\mathcal F_0+\int_{\bq}\eta_I(-\bq)[q_\mu B_{\mu I}(\bq)+\ii Q_{{\rm eff},I}]$ gives $K^{(s)}_{\mu\nu}\Xi_{IJ}B_{\nu J}+q_\mu\eta_I=0$, so that
\begin{equation}
\label{eq:supp-lambda-solution}
 \eta_I(\bq)={\ii\Xi_{IJ}Q_{{\rm eff},J}\over\bq\cdot\mathsf A\bq},
\end{equation}
and
\begin{equation}
\label{eq:supp-strict-fourier-solution}
 B_{\mu I}(\bq)=-\ii Q_{{\rm eff},I}{(\mathsf A\bq)_\mu\over\bq\cdot\mathsf A\bq}.
\end{equation}
The spectator metric cancels because the same Gauss law is imposed in every spectator sector.  With the anisotropic Green function $\int_{\bq}e^{\ii\bq\cdot\br}/\bq\cdot\mathsf A\bq=-\ln(r_\mu\mathsf A^{-1}_{\mu\nu}r_\nu)/(4\pi\sqrt{\det\mathsf A})+{\rm const}$,
\begin{equation}
\label{eq:supp-anisotropic-real-space}
 B_{\mu I}(\br)=
 {Q_{{\rm eff},I}\sqrt{\det K^{(s)}}\over2\pi}
 {r_\mu\over r_\rho K^{(s)}_{\rho\sigma}r_\sigma}.
\end{equation}
The denominator is positive, so anisotropy changes only the amplitude: $b_\mu=v_IB_{\mu I}$, for any $v_I$ with $v_IQ_{{\rm eff},I}\ne0$, is radial and winds once, and the soft modes $\Phi_A=(\mathcal M^{-T})_{A\mu}b_\mu$ wind $\sgn\det\mathcal M=W_{\rm bulk}$ times.

\paragraph*{Screened monopole.}
For $\textbf d=\ii\mathcal M\bk$ with $\mathcal M^T\mathcal M=\lambda\mathbb I$, Eq.~\eqref{eq:supp-hybrid-continuum} gives $\delta\langle\textbf B(\bk)\rangle=-\ii Q_{\rm eff}\lambda\bk/(\kappa T+\lambda k^2)$.  The angular integral is the Lemma of Sec.~S4 with $m=\pm1$, leaving a single Hankel transform,
\begin{equation}
\label{eq:supp-rank1-hybrid-texture}
 \delta\langle\bar B_\mu(\br)\rangle=\frac{Q_{\rm eff}}{2\pi}\,
 \frac{r_\mu}{r}\int_0^\infty\!\dd k\,
 \frac{k^2\,J_1(kr)}{k^2+m^2}
 =\frac{Q_{\rm eff}}{2\pi}\,
 m K_1(mr)\,\frac{r_\mu}{r},
 \qquad m=\xi_1^{-1}=\sqrt{\kappa T/\lambda},
\end{equation}
with induced charge density
\begin{equation}
\label{eq:supp-screening-cloud}
 \delta\langle\rho(\br)\rangle
 =Q_{\rm eff}\left[\delta^{(2)}(\br)-\frac{m^2}{2\pi}K_0(mr)\right].
\end{equation}
The two terms integrate to zero: the orphan's point charge is neutralized by a screening cloud of radius $\xi_1$, and a disc of radius $R$ encloses the charge $Q_{\rm eff}(R/\xi_1)K_1(R/\xi_1)$.  In the Coulomb window $mK_1(mr)\simeq1/r$, so the texture is the strict monopole, radial with unit winding.  Because $m\propto T^{1/2}$, the correlator obeys $\mathcal D\propto T^{3/2}f_1(rT^{1/2})$, the standard rank-1 Coulomb-phase scaling, while the texture is $\delta\langle\bar{\textbf B}\rangle\propto Q_{\rm eff}T^{1/2}f_1(rT^{1/2})$: the factor $T$ in $\mathcal D$ is removed by $1/\sigma_0^2$.

\subsection*{S6. Rank 2: tensor gauge field}\phantomsection\label{sec:S6}

At $n=2$ the gauge field is a symmetric traceless tensor with $B_{xx}=-B_{yy}=B_1$ and $B_{xy}=B_{yx}=B_2$, and its Gauss law has the canonical symbol $\textbf d_0(\bk)=(k_x^2-k_y^2,2k_xk_y)$, of degree $+2$; Sec.~S7 shows that the honeycomb-snowflake model realizes it with $\textbf B=-\sqrt\lambda\,\bm\Phi$.

\paragraph*{Point charge.}
When the charged field is a valley amplitude, as for the $\bm K$ valley of Sec.~S7, its real and imaginary parts are identical real sectors, so it is enough to solve one.  With the differential operator $\textbf d_0(\bm\partial)$ the Gauss law reads
\begin{equation}
\label{eq:supp-rank2-D-operator}
 \textbf d_0(\bm\partial)=(\partial_x^2-\partial_y^2,2\partial_x\partial_y),
 \qquad
 \partial_\mu\partial_\nu B_{\mu\nu}=d_{0,A}(\bm\partial)B_A=Q\delta^{(2)}(\br).
\end{equation}
Minimizing the isotropic functional $\mathcal F_2=(\kappa/2)\int\dd^2r\,B_AB_A$ subject to this charge gives
\begin{equation}
\label{eq:supp-rank2-exact-saddle}
 B_A(\bk)=-Q\frac{d_{0,A}(\bk)}{k^4},
 \qquad
 B_A(\br)=Q\,d_{0,A}(\bm\partial)G_4(\br)
 =\frac{Q}{4\pi}(\cos2\theta,\sin2\theta)_A,
\end{equation}
for $r>0$, where $G_4(\br)=r^2\ln(r/a)/(8\pi)$ obeys $\nabla^4G_4=\delta^{(2)}(\br)$ and $d_{0,A}(\bm\partial)d_{0,A}(\bm\partial)=\nabla^4$.  This is the $n=2$ case of Eq.~\eqref{eq:supp-canonical-point-charge}, and shows compactly why a rank-2 charge produces a nondecaying quadrupolar field rather than the $1/r$ field of rank 1.

\paragraph*{Screened texture.}
With the soft modes normalized as in Sec.~S2 and $\textbf d=\sqrt\lambda\,\textbf d_0$, $d_1+\ii d_2=\sqrt\lambda\,k^2e^{2\ii\phi}$ and Eq.~\eqref{eq:supp-hybrid-continuum} carries the single angular harmonic $e^{2\ii\phi}$.  The Lemma of Sec.~S4 with $m=2$ gives
\begin{equation}
\label{eq:supp-rank2-texture}
 \begin{aligned}
 \delta\langle \Phi^z_c(\br)\rangle
 &=-\frac{Q_{\rm eff}e^{2\ii\theta}}{2\pi}\int_0^\infty\!\dd k\,
 \frac{\sqrt\lambda\,k^3J_2(kr)}{\kappa T+\lambda k^4}
 =-\frac{Q_{\rm eff}}{4\pi\sqrt\lambda}\,F_2(r/\xi_2)\,e^{2\ii\theta},\\
 F_2(x)&=2\int_0^\infty\!\dd u\,\frac{u^3J_2(u)}{u^4+x^4},
 \end{aligned}
\end{equation}
with $\xi_2=(\lambda/\kappa T)^{1/4}$ and $F_2\to1$ for $r\ll\xi_2$, since $\mathcal I(2,2)=1/2$ in Eq.~\eqref{eq:supp-radial-multiplier}.  For the gauge field this is $\delta\langle B^z_c(\br)\rangle=(Q_{\rm eff}/4\pi)F_2(r/\xi_2)\,e^{2\ii\theta}$: in the Coulomb window the nondecaying point-charge field of Eq.~\eqref{eq:supp-rank2-exact-saddle} with $Q=Q_{\rm eff}$, screened beyond $\xi_2\propto T^{-1/4}$.  Integrating the induced charge density $Q_{\rm eff}\lambda k^4/(\kappa T+\lambda k^4)$ over a disc of radius $R$ gives the enclosed charge
\begin{equation}
\label{eq:supp-rank2-G2}
 Q_{\rm eff}\,G_2\!\left(\frac R{\xi_2}\right),
 \qquad
 G_2(x)=1-x\int_0^\infty\!\dd u\,\frac{J_1(ux)}{1+u^4},
\end{equation}
a point charge at short distance and a neutralizing cloud beyond $\xi_2$; unlike its rank-1 counterpart the cloud overscreens, $G_2$ crossing zero at $x=2.7$ with a minimum of $-0.13$ at $x=3.9$.

\subsection*{S7. Microscopic models}\phantomsection\label{sec:S7}

\begin{table}[t!]
\centering
\small
\setlength{\tabcolsep}{5pt}
\begin{tabular}{llllll}
\hline\hline
model & $\bq_0$ & soft constrainer & $v(\bq_0)$ & measured charge & charged field \\
\hline
kagome & $\Gamma$ & $(M_{\mathcal A}-M_{\mathcal B})/\sqrt2$ & $1/2$ & staggered vertex charge $\eta_vM_v$ & $\textbf B_{\rm K}=\frac23\sum_a\hat{\bm e}_aS_a$ \\
checkerboard & $(\pi,\pi)$ & vertex constraint $M$ & $1$ & staggered vertex charge $\eta_{\bR}M$ & $\textbf B=\eta_{\bR}(S_x,S_y)$ \\
snowflake & $\pm\bm K$ & snowflake constraint $C_{\hexagon}$ & $1$ & valley charge of $\delta C_{\hexagon}$ & $\Phi_c=\sqrt2\,\psi_1$, $\textbf B=-\sqrt\lambda\,\bm\Phi$ \\
\hline\hline
\end{tabular}
\caption{Normalizations of the three models.  $v(\bq_0)=|U_{l_0s}(\bq_0)|^2$ is the overlap of the orphan constrainer with the soft band, which enters Eq.~\eqref{eq:supp-sigma0-value}.  On kagome the orphan simplex couples to the normalized soft combination with amplitude $1/\sqrt2$, and the staggered charge of a cell is $\sqrt2$ times that combination, so in the units of the measured charge the point charge is $Q_{\rm eff}=T\langle\mu^z\rangle$ (one source) on all three lattices.  Continuum soft modes are densities per unit-cell area, $A_{\rm uc}=3\sqrt3/2$, $1$ and $\sqrt3/2$ in the units of Sec.~S8; the charged fields are those of Eqs.~\eqref{eq:supp-kagome-flux-reconstruction}, \eqref{eq:supp-checkerboard-flux-reconstruction} and \eqref{eq:supp-rank2-field-reconstruction}.}
\label{tab:models}
\end{table}

\paragraph*{Line-graph formulation.}
Let $P$ be the bipartite parent graph: honeycomb for kagome and square for checkerboard.  A unit Heisenberg spin $\vec S_\ell$ lives on each active parent link $\ell$, and two spins interact when their links share a parent vertex.  The Hamiltonian of Eq.~\eqref{eq:kc-hamiltonian-prl} of the main text can be written exactly as
\begin{align}
\label{eq:supp-line-graph-hamiltonian}
  H
  &=J\sum_{\langle\ell\ell'\rangle_{L(P)}}
  \vec S_\ell\cdot\vec S_{\ell'}
  -\sum_{\ell}h_\ell S_\ell^z\\
  &={J\over2}\sum_{v\in P}|\vec M_v|^2
  -JN_{\rm act}-\sum_\ell h_\ell S_\ell^z,
  \qquad
  \vec M_v\equiv\sum_{\substack{\ell\ni v\\ \ell\ {\rm active}}}\vec S_\ell .
\end{align}
Thus the low-energy constraints are $\vec M_v=0$ wherever the active local geometry permits them.  Orient every link from parent sublattice $\mathcal A$ to $\mathcal B$, take $\eta_v=+1$ on $\mathcal A$ and $-1$ on $\mathcal B$, and define
\begin{equation}
\label{eq:supp-general-incidence-charge}
  \epsilon_{v\ell}=\eta_v\,\mathbf 1_{v\in\partial\ell},
  \qquad
  \vec Q_v\equiv\eta_v\vec M_v
  =\sum_\ell\epsilon_{v\ell}\vec S_\ell .
\end{equation}
Equation~\eqref{eq:supp-general-incidence-charge} distinguishes the physical simplex magnetization $\vec M_v$ from the staggered gauge charge $\vec Q_v$.  On an oriented link with unit tangent $\hat{\bm e}_\ell$, the microscopic flux is $\bm b_\ell^I=\hat{\bm e}_\ell S_\ell^I$, and its lattice divergence is $Q_v^I$ up to the fixed cell-volume normalization.

\paragraph*{Kagome and checkerboard.}
Choose honeycomb Bravais vectors and the three oriented $\mathcal A$-to-$\mathcal B$ link directions
\begin{align}
\label{eq:supp-kagome-geometry}
  \bm a_1&=(\sqrt3,0),&
  \bm a_2&=(\sqrt3/2,3/2),\\
  \hat{\bm e}_0&=(0,1),&
  \hat{\bm e}_1&=(-\sqrt3/2,-1/2),&
  \hat{\bm e}_2&=(\sqrt3/2,-1/2).
\end{align}
The links $a=0,1,2$ beginning at $\mathcal A(\bR)$ end at $\mathcal B(\bR)$, $\mathcal B(\bR-\bm a_2)$, and $\mathcal B(\bR+\bm a_1-\bm a_2)$, respectively.  With $\vec S(\bq)=(\vec S_0,\vec S_1,\vec S_2)^T$ and cell-based Fourier phases, the two physical simplex constraints are
\begin{equation}
\label{eq:supp-kagome-constraint-matrix}
  \begin{pmatrix}\vec M_{\mathcal A}(\bq)\\ \vec M_{\mathcal B}(\bq)\end{pmatrix}
  =\mathsf L_{\rm K}(\bq)\vec S(\bq),
  \qquad
  \mathsf L_{\rm K}(\bq)=
  \begin{pmatrix}
    1&1&1\\
    1&e^{\ii\bq\cdot\bm a_2}&e^{\ii\bq\cdot(\bm a_2-\bm a_1)}
  \end{pmatrix}.
\end{equation}
At $\bq_0=\bm0$ the two rows of $\mathsf L_{\rm K}$ coincide and its rank drops by one.  The row combination that vanishes at the pinch point is
\begin{align}
\label{eq:supp-kagome-small-q-row}
  \bm C_{\rm K}(\bq)
  &\equiv(\mathsf L_{{\rm K},\mathcal B}-\mathsf L_{{\rm K},\mathcal A})(\bq)\\
  &=\ii\left(0,\bq\cdot\bm a_2,
  \bq\cdot(\bm a_2-\bm a_1)\right)+O(q^2).
\end{align}
On the $\mathcal A$-simplex constraint subspace $\sum_aS_a^I=0$, define the two-component flux
\begin{equation}
\label{eq:supp-kagome-flux-reconstruction}
  \textbf B_{\rm K}^I={2\over3}\sum_{a=0}^2
  \hat{\bm e}_a S_a^I,
  \qquad
  S_a^I=\hat{\bm e}_a\cdot\textbf B_{\rm K}^I.
\end{equation}
The inverse follows from $\sum_a\hat{\bm e}_a=0$ and
$\sum_a\hat e_{a\mu}\hat e_{a\nu}=(3/2)\delta_{\mu\nu}$, so the reconstruction is nonsingular on the charged rank-1 soft sector.  Substitution into Eq.~\eqref{eq:supp-kagome-small-q-row} gives
\begin{equation}
\label{eq:supp-kagome-principal-symbol}
  \bm C_{\rm K}(\bq)\cdot\vec S(\bq)
  =-{3\ii\over2}\,\bq\cdot\textbf B_{\rm K}+O(q^2).
\end{equation}
Thus the kagome pinch point is at $\Gamma$ (modulo reciprocal vectors), its spatial principal matrix is $-(3/2)\mathbb I$, and its degree is $+1$ with the chosen coordinate orientation.

The kagome vacancy cluster is placed on a central $\mathcal A$ vertex $\bR_0$.  Of its three incident link spins, the $a=0$ spin is retained and the $a=1,2$ spins are removed.  Hence
\begin{equation}
\label{eq:supp-kagome-defect-charge}
  \vec M_{\mathcal A,0}=\vec S_0,
  \qquad \vec Q_{\mathcal A,0}=\vec S_0,
\end{equation}
so $m_{\rm O}=1$.  The other simplices touched by a removed spin retain two active spins and can satisfy their constraint by antiparallel alignment; only the one-spin central simplex cannot, so the polarized survivor is a pure monopole source, $Q^z(\bk)=\langle S_0^z\rangle+O(k)$.

Let $\bR=(R_x,R_y)\in\mathbb Z^2$ label square-parent vertices, and let $S_x^I(\bR)$ and $S_y^I(\bR)$ live on the links from $\bR$ in the positive $x$ and $y$ directions.  The physical simplex constraint is
\begin{equation}
\label{eq:supp-checkerboard-real-constraint}
  M^I(\bR)=S_x^I(\bR)+S_x^I(\bR-\hat{\bm x})
  +S_y^I(\bR)+S_y^I(\bR-\hat{\bm y}).
\end{equation}
In the cell-based Fourier convention,
\begin{equation}
\label{eq:supp-checkerboard-constraint-vector}
  M^I(\bq)=\bm C_{\rm CB}(\bq)\cdot
  \begin{pmatrix}S_x^I(\bq)\\S_y^I(\bq)\end{pmatrix},
  \qquad
  \bm C_{\rm CB}(\bq)=
  \left(1+e^{-\ii q_x},1+e^{-\ii q_y}\right).
\end{equation}
Harmless link-midpoint phases convert these entries to $2\cos(q_x/2)$ and $2\cos(q_y/2)$.  The raw-spin pinch point is
$\bq_0=(\pi,\pi)$, and for $\bq=\bq_0+\bk$,
\begin{equation}
\label{eq:supp-checkerboard-small-q}
  \bm C_{\rm CB}(\bq_0+\bk)=\ii(k_x,k_y)+O(k^2).
\end{equation}
Equivalently, set $\eta_{\bR}=(-1)^{R_x+R_y}$ and reconstruct
\begin{equation}
\label{eq:supp-checkerboard-flux-reconstruction}
  B_x^I(\bR)=\eta_{\bR}S_x^I(\bR),
  \qquad B_y^I(\bR)=\eta_{\bR}S_y^I(\bR).
\end{equation}
This map is exactly invertible.  The staggered charge becomes the square-lattice divergence
\begin{align}
\label{eq:supp-checkerboard-divergence}
  Q^I(\bR)=\eta_{\bR}M^I(\bR)
  &=B_x^I(\bR)-B_x^I(\bR-\hat{\bm x})\\
  &\quad+B_y^I(\bR)-B_y^I(\bR-\hat{\bm y}),
\end{align}
whose symbol at flux momentum $\bk=\bm0$ is
\begin{equation}
\label{eq:supp-checkerboard-principal-symbol}
  \textbf d_{\rm CB}(\bk)=
  \left(1-e^{-\ii k_x},1-e^{-\ii k_y}\right)
  =\ii(k_x,k_y)+O(k^2).
\end{equation}
The checkerboard principal spatial matrix is therefore the identity and has winding $+1$.  The apparent difference between $\bq_0=(\pi,\pi)$ in raw spins and $\bk=0$ in the flux field is exactly the staggering in Eq.~\eqref{eq:supp-checkerboard-flux-reconstruction}.

For the checkerboard vacancy cluster, choose a central even-parity parent vertex $\bR_0$, retain the $+x$ link spin $S_x(\bR_0)$, and remove the other three incident spins.  Then
\begin{equation}
\label{eq:supp-checkerboard-defect-charge}
  \vec M_0=\vec S_x(\bR_0),
  \qquad \vec Q_0=\vec S_x(\bR_0),
\end{equation}
so again $m_{\rm O}=1$; every other simplex touched by a removed spin retains three active spins and can satisfy its constraint, and the polarized survivor is a pure monopole source.

\paragraph*{Honeycomb-snowflake model.}
The model of Eq.~\eqref{eq:rank2-hamiltonian-prl} of the main text and its higher-rank point were introduced in Ref.~\cite{BentonMoessner21}; the constrainer expansion was worked out systematically in Ref.~\cite{YanClassification}.  Each snowflake constrainer is $\vec C_{\hexagon}=\sum_ic_{\hexagon i}\vec S_i$ with $c_{\hexagon i}=1$ on the six sites of hexagon $\hexagon$ and $\gamma$ on its six exterior neighbours.  We take Bravais vectors $\bm a_1=(1,0)$ and $\bm a_2=(1/2,\sqrt3/2)$, sublattice positions $\bm\tau_1=(1/2,\sqrt3/6)$ and $\bm\tau_2=(1,1/\sqrt3)$, so that hexagon $(i,j)$ is centred at $i\bm a_1+j\bm a_2$, and $\bm K=(4\pi/3,0)$.  With phases measured from the hexagon centre, direct expansion of the twelve site phases at $\gamma=1/2$ gives
\begin{equation}
\label{eq:supp-rank2-constrainer-expansion}
 \big(T^1,T^2\big)(\bm K+\bk)
 =\frac38\left(k_+^2,k_-^2\right)+O(k^3),
 \qquad k_\pm=k_x\pm\ii k_y,
\end{equation}
so that, with $\vec\psi_a(\bk)\equiv\vec S^a(\bm K+\bk)$, the constrainer is $\sum_aT^a(\bm K+\bk)^*\vec\psi_a=\frac38(k_-^2\vec\psi_1+k_+^2\vec\psi_2)$ and the dispersive band is $\omega(\bm K+\bk)=\lambda k^4$ with $\lambda=9/32$.  The coefficient reflects the choice of lattice spacing; the quadratic zero and its degree do not.  Reality of the spins fixes the expansion at $-\bm K$ by complex conjugation.  The soft modes and the gauge field $\textbf B=-\sqrt\lambda\,\bm\Phi$ (Sec.~S3) are
\begin{equation}
\label{eq:supp-rank2-field-reconstruction}
 \Phi_1^I=\frac{\psi_1^I+\psi_2^I}{\sqrt2},
 \qquad
 \Phi_2^I=\ii\,\frac{\psi_2^I-\psi_1^I}{\sqrt2},
 \qquad
 \mathsf B^I=\begin{pmatrix}B_1^I&B_2^I\\B_2^I&-B_1^I\end{pmatrix}
 =-\sqrt\lambda\begin{pmatrix}\Phi_1^I&\Phi_2^I\\\Phi_2^I&-\Phi_1^I\end{pmatrix},
\end{equation}
with $\sqrt\lambda=3\sqrt2/8$.
Substitution into Eq.~\eqref{eq:supp-rank2-constrainer-expansion} yields
\begin{equation}
\label{eq:supp-rank2-gauss-law}
 C^I(\bm K+\bk)
 =d_A(\bk)\Phi_A^I(\bk)=-d_{0,A}(\bk)B_A^I(\bk)+O(k^3),
 \quad
 \textbf d=\sqrt\lambda\,\textbf d_0,
 \quad
 \partial_\mu\partial_\nu B_{\mu\nu}^I=0 .
\end{equation}
The last equality is the real-space form of the ground-state constraint: $-\textbf d_0(\bk)\cdot\textbf B$ is the Fourier transform of $\partial_\mu\partial_\nu B_{\mu\nu}$, the sign being the $\ii^2$ of Eq.~\eqref{eq:supp-canonical-gauss} that $L=-\sqrt\lambda\,\mathbb I$ compensates.  On $\bk=k(\cos\phi,\sin\phi)$, $\hat{\textbf d}=(\cos2\phi,\sin2\phi)$ has degree $+2=2\,\sgn\det L$.  Notice also that
\begin{equation}
\label{eq:supp-rank2-chiral-fields}
 \Phi_1^I+\ii\Phi_2^I=\sqrt2\,\psi_1^I,
 \qquad
 \Phi_1^I-\ii\Phi_2^I=\sqrt2\,\psi_2^I.
\end{equation}
The two sublattices therefore carry opposite chiral projections of the same traceless tensor.

The vacancy clusters of Ref.~\cite{FloresCalderonBentonMoessner24} keep a single spin in the snowflake of one hexagon $\hexagon_0$: cluster O1 removes the six exterior sites and five of the six hexagon sites, keeping a hexagon site, so $m_{\rm O}=1$; cluster O2 keeps instead an exterior site, so $m_{\rm O}=\gamma$.  In both cases eleven sites are removed and $\vec C_{\hexagon_0}=m_{\rm O}\vec S_{\rm O}$.

By Eq.~\eqref{eq:supp-rank2-chiral-fields} and the density normalization of Sec.~S2, Eq.~\eqref{eq:supp-rank2-texture} and its complex conjugate at $-\bm K$ give the sublattice textures
\begin{equation}
\label{eq:supp-rank2-derived-texture}
 \delta\langle S^{a,z}(\br)\rangle=\frac{2A_{\rm uc}Q_{\rm eff}}{3\pi}\,F_2(r/\xi_2)\,
 \cos\!\left(\bm K\cdot\br+2\eta_a\theta+\chi\right),
 \qquad \eta_1=+1,\quad\eta_2=-1,
\end{equation}
with $\br$ measured from the centre of $\hexagon_0$ and $\chi$ a constant phase.  We checked Eq.~\eqref{eq:supp-rank2-derived-texture} against a direct lattice evaluation of Eq.~\eqref{eq:texture-factorization-prl} of the main text ($300\times300$ momentum grid, $T=3\times10^{-6}J$, self-consistent $\kappa=1.50$, $\xi_2=15.8$): the demodulated $+\bm K$ amplitude at $R=5$--$7$ agrees with $A_{\rm uc}Q_{\rm eff}F_2/3\pi$ to within $2\%$, and its phase follows $-e^{2\ii\theta}$.
\subsection*{S8. Monte Carlo simulations}\phantomsection\label{sec:S8}

This section gives the protocol and results behind Figs.~\ref{fig:rank1-mc-winding} and \ref{fig:rank2-mc-winding} of the main text for all three lattices.  Each test addresses the three claims of the main text: the defect winding equals $W_{\rm bulk}$ on every circle of the Coulomb window, the amplitude of the texture is predicted without free parameters from the charge $Q_{\rm core}=m_{\rm O}\langle n^z\rangle-\langle C^z_0\rangle_{{\rm ud},h}$ imposed on the orphan, and the screening length scales as $\xi_n\propto T^{-1/2n}$.

\begin{figure}[t!]
\centering
\includegraphics[width=\textwidth]{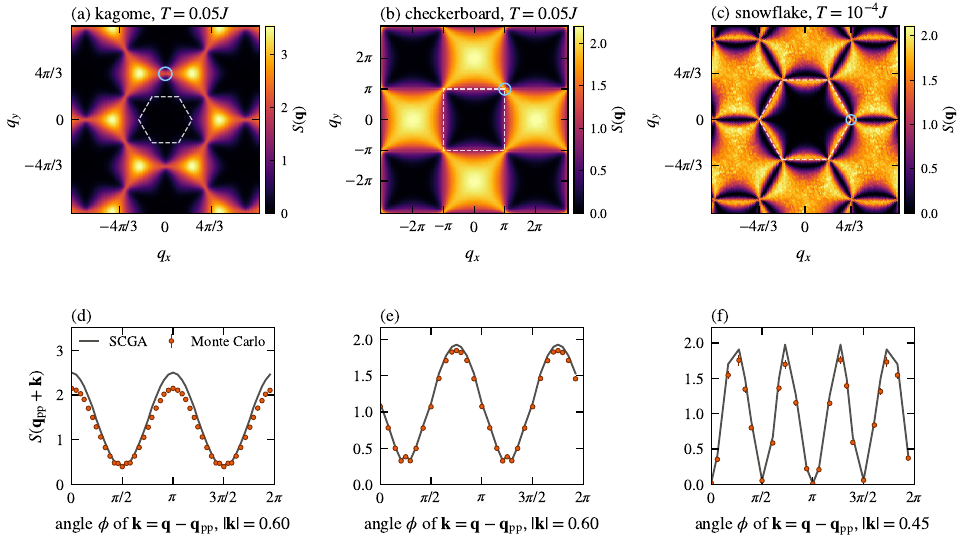}
\caption{Clean structure factor at the simulated parameters ($L=48$, $h=0.2J$).  (a)--(c)~Connected $S(\bq)$ from Monte Carlo on the extended zone; dashed: first Brillouin zone; circle: ring around the pinch point $\bq_{\rm pp}=(0,4\pi/3)$, $(\pi,\pi)$ and $\bm K=(4\pi/3,0)$ used in (d)--(f).  Wavevectors are in units of the inverse parent bond length at rank 1 and of the inverse Bravais spacing at rank 2.  (d)--(f)~$S$ on that ring against the angle of $\bk=\bq-\bq_{\rm pp}$: Monte Carlo (points, standard errors over blocks) and the SCGA at the same momenta (lines).  The two-fold modulation at rank 1 and the four-fold one at rank 2 are the pinch points of the Coulomb and rank-2 liquids.}
\label{fig:S-structure-factor}
\end{figure}
\paragraph*{Models, sampling and spin-liquid window.}
We simulate the classical Heisenberg models of Eqs.~\eqref{eq:kc-hamiltonian-prl} and \eqref{eq:rank2-hamiltonian-prl} of the main text, $|\vec S_i|=1$, in the uniform field $h=0.2J$, with the vacancy clusters of Sec.~S7 on periodic $L\times L$ arrays of parent cells ($3L^2$ spins on kagome, $2L^2$ on checkerboard and honeycomb, with $3\,|\,L$ for the latter so that $\pm\bm K$ are lattice momenta).  Lengths are in units of the parent bond length at rank 1 and of the Bravais spacing at rank 2 (Sec.~S7).  One compound sweep is a heat-bath sweep in random order, in which every spin is redrawn from $P(\vec S)\propto\exp(\beta\,\vec b\cdot\vec S)$ in its local field $\vec b$, followed by five overrelaxation sweeps, $\vec S\to2(\vec S\cdot\vec b)\vec b/b^2-\vec S$.  Snowflake chains are first annealed in 20 stages from $\beta J=0.1$ to $J/T$ with 250 sweeps per stage.  After equilibration ($5\times10^3$ sweeps, $10^4$ at $T=0.025J$) each chain records 40 blocks of $2.5\times10^3$ sweeps; we run four chains per data set (eight at rank 1 for $L=48$, $T=0.05J$, and sixteen for the main rank-2 data set and for $L=36$ at $T=10^{-4}J$) and, for every $(L,T,h)$, a reference run of two chains without vacancies.  The energy agrees between the first and last quarter of the blocks to $10^{-5}J$ per link (rank 1) and $2\times10^{-7}J$ per site (rank 2).  The data sets are listed in Tables~\ref{tab:mc-summary} and \ref{tab:mc-rank2}; at rank 2 we also ran $L=36$ at $T=10^{-4}J$ and $3\times10^{-4}J$ for the finite-size check.  At rank 2 the integrated autocorrelation time of the texture reaches $4\times10^3$ sweeps at $T=10^{-4}J$, still much shorter than the jackknife groups below, and exceeds $10^4$ sweeps at $10^{-5}J$, which restricts production to $T\ge10^{-4}J$.

To confirm that the simulations probe the spin liquid, we measured the connected structure factor $S(\bq)=N^{-1}\sum_{ij}\langle\vec S_i\cdot\vec S_j\rangle_{\rm c}\,e^{\ii\bq\cdot(\br_i-\br_j)}$ of the undiluted lattices at the production parameters, $L=48$ and $h=0.2J$, in separate runs with the reference protocol, four chains of $5\times10^4$ sweeps at rank 1 and eight annealed chains of $10^5$ sweeps at rank 2, recording $S(\bq)$ every 10 sweeps (Fig.~\ref{fig:S-structure-factor}); they reproduce the magnetizations $m_h$ of the reference runs.  All three lattices show the pinch points of the band touching, two-fold on kagome and checkerboard and four-fold on the snowflake lattice, whose positions, widths and angular structure follow the SCGA [Fig.~\ref{fig:S-structure-factor}(d)--(f)], to within $4\%$ on checkerboard.  On kagome and on the snowflake lattice the Monte Carlo lobes are $10$--$15\%$ weaker than the SCGA, the weight moving to broad maxima, on kagome at the corners of the extended zone, $|\bq|=8\pi/3\sqrt3$, where the classical kagome magnet develops short-range $\sqrt3\times\sqrt3$ correlations that a Gaussian theory omits.  No Bragg peak develops: the largest value of $S(\bq)$ over the extended zones, $3.6$ (kagome), $2.2$ (checkerboard) and $2.4$ (snowflake), does not grow with $L$ between $L=24$ (kagome), $32$ (checkerboard) or $36$ (snowflake) and $L=48$, so that $S_{\max}/N\approx5\times10^{-4}$ at $L=48$, whereas long-range order would give $S/N=O(1)$; the only extensive weight is the field-induced magnetization in the uniform mode, which the connected subtraction removes.  The same holds at $T=0.025J$, with $S_{\max}=4.7$ on kagome, also at $L=32$, and $2.4$ on checkerboard.

\begin{figure}[t!]
\centering
\includegraphics[width=\textwidth]{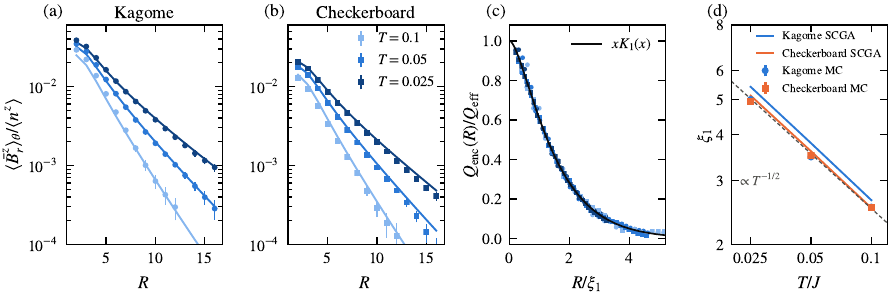}
\caption{Temperature dependence and screening ($L=48$, uniform field $h=0.2J$).  (a,b) Angular average of the radial flux per unit orphan magnetization for $T/J=0.1$, $0.05$, $0.025$ (light to dark); points are Monte Carlo data with jackknife errors, lines the SCGA prediction of Eq.~\eqref{eq:supp-lattice-saddle} in the same field.  (c) Enclosed gauge charge, Eq.~\eqref{eq:supp-enclosed-charge}, divided by its fitted short-distance value $Q_{\rm eff}$ and plotted against $R/\xi_1$, for both lattices (circles: kagome, squares: checkerboard) and all three temperatures; the line is $xK_1(x)$.  (d) Screening length from the fits in (c), with the SCGA values (lines) and a $T^{-1/2}$ guide (dashed).}
\label{fig:S-scaling}
\end{figure}

\begin{table}[b!]
\centering
\small
\setlength{\tabcolsep}{6pt}
\begin{tabular}{lllllllll}
\hline\hline
lattice & $L$ & $T/J$ & $\alpha$ & $W_{\rm MC}=1$ for & $Q_{\rm eff}$ & $Q_{\rm core}$ & $Q_{\rm eff}^{\rm SCGA}$ & $\xi_1$ \\
\hline
kagome & 48 & 0.1 & 0.531 & $R=3$--$6$ & $0.291(6)$ & 0.226 & 0.235 & $2.53(8)$ \\
kagome & 48 & 0.05 & 0.517 & $R=2$--$11$ & $0.519(2)$ & 0.446 & 0.456 & $3.48(3)$ \\
kagome & 48 & 0.025 & 0.510 & $R=2$--$14$ & $0.717(3)$ & 0.654 & 0.662 & $5.00(9)$ \\
kagome & 32 & 0.05 & 0.517 & $R=2$--$10^{\,*}$ & $0.515(4)$ & 0.447 & 0.455 & $3.53(6)$ \\
checkerboard & 48 & 0.1 & 0.526 & $R=4$--$7$ & $0.209(4)$ & 0.232 & 0.240 & $2.52(10)$ \\
checkerboard & 48 & 0.05 & 0.513 & $R=2$--$12$ & $0.445(3)$ & 0.449 & 0.458 & $3.51(6)$ \\
checkerboard & 48 & 0.025 & 0.505 & $R=2$--$13$ & $0.654(3)$ & 0.655 & 0.662 & $4.95(7)$ \\
checkerboard & 32 & 0.05 & 0.515 & $R=2$--$9$ & $0.441(3)$ & 0.451 & 0.460 & $3.53(6)$ \\
\hline\hline
\end{tabular}
\caption{Rank-1 data sets (uniform field $h=0.2J$).  $\alpha$ is the renormalized orphan moment, from $\langle n^z\rangle=L(\alpha h/T)$; the winding column lists the radii with a robust $W_{\rm MC}=1$ (asterisk: the full range $R\le L/3$).  $Q_{\rm eff}$ and $\xi_1$ come from fits of Eq.~\eqref{eq:supp-enclosed-charge}, with jackknife errors of the refits; $Q_{\rm core}=\langle n^z\rangle-zm_h$ is the charge imposed on the orphan, and $Q_{\rm eff}^{\rm SCGA}$ the result of the same fit applied to the lattice SCGA texture of Eq.~\eqref{eq:supp-lattice-saddle}.}
\label{tab:mc-summary}
\end{table}

\paragraph*{Analysis and orphan-field protocol.}
The field polarizes the undiluted liquid uniformly, $\langle S^z_i\rangle_{{\rm ref},h}=m_h$, and we analyse $\delta S^z_i=\langle S^z_i\rangle_{{\rm defect},h}-m_h$ per unit orphan magnetization $\langle n^z\rangle=\langle S^z_{\rm O}\rangle$, measured in the same run.  At $T=0.05J$, $m_h=0.0342$ (kagome) and $0.0241$ (checkerboard), close to $h/6J$ and $h/8J$; on the snowflake lattice $m_h=0.004938=2h/81J$ at every $(L,T)$, equal to the SCGA to $10^{-6}$.  The orphan constrainer carries the charge $Q_{\rm core}=m_{\rm O}\langle n^z\rangle-\langle C^z_0\rangle_{{\rm ud},h}$ of Sec.~S2, with $\langle C^z_0\rangle_{{\rm ud},h}=zm_h$ for $z=3$ (kagome) or $4$ (checkerboard) links per vertex, and $\langle C^z_0\rangle_{{\rm ud},h}=(6+6\gamma)m_h=2h/9J=0.044$ at rank 2.  As a cross-check we also simulated a field on the orphan alone, $-h_{\rm O}S^z_{\rm O}$ with $h_{\rm O}=2J$, for every data set.  The rest of the system is then invariant under global spin rotations: the mean of every spin at fixed orphan direction is $\langle\vec S_i\rangle_{\vec n}=f_i\,\vec n$, and the orphan's distribution is $P(\vec n)\propto e^{\beta h_{\rm O}n^z}$, so that without approximation
\begin{equation}
\label{eq:supp-exact-factorization}
 \delta\langle S_i^z\rangle=f_i\,L(\beta h_{\rm O}),
 \qquad
 f_i=\langle\vec S_i\cdot\vec S_{\rm O}\rangle,
\end{equation}
Eq.~\eqref{eq:texture-factorization-prl} of the main text with $\langle C^z_0\rangle_{{\rm ud},h}=0$ and no reference subtraction.  The two sides agree link by link to $10^{-3}$ (rank 1) and site by site to $3\times10^{-4}$ (rank 2).

\begin{figure}[t!]
\centering
\includegraphics[width=0.68\textwidth]{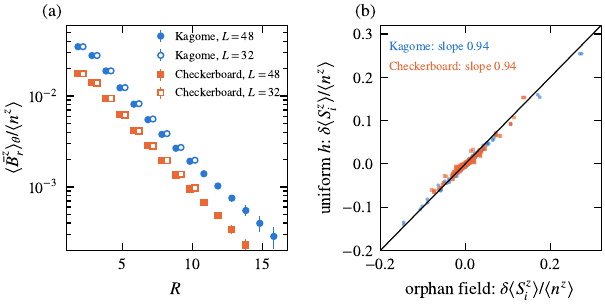}
\caption{Rank-1 checks ($T=0.05J$, uniform field $h=0.2J$).  (a) Radial flux per unit orphan magnetization for $L=48$ (filled) and $L=32$ (open); circles: kagome, squares: checkerboard.  (b) Texture per unit orphan magnetization, $\delta\langle S^z_i\rangle/\langle n^z\rangle$, in the uniform field (vertical) against the orphan-field protocol of Eq.~\eqref{eq:supp-exact-factorization} (horizontal), for every active link within $R\le6$ of the orphan, the orphan link excluded ($L=48$); the line is the identity, and the slopes are those of fits through the origin.}
\label{fig:S-checks}
\end{figure}

At rank 1 the link flux $b^z_{\ell\mu}=\hat e_{\ell\mu}\,\delta S^z_\ell$, the real-space form of Eqs.~\eqref{eq:supp-kagome-flux-reconstruction} and \eqref{eq:supp-checkerboard-flux-reconstruction}, is coarse grained,
\begin{equation}
\label{eq:supp-coarse-flux}
  \bar B_\mu^z(\bR)
  =
  {\sum_{\ell} w_\sigma(\bR-\br_\ell)b_{\ell\mu}^z
  \over
  \sum_{\ell} w_\sigma(\bR-\br_\ell)} ,
  \qquad
  w_\sigma(\br)=e^{-r^2/2\sigma^2}\,\Theta(4\sigma-r),
\end{equation}
with $\sigma=1.4$, the sums running over active links with the minimum image.  A slowly varying flux density gives $\delta S^z_\ell=w_\ell\,\hat e_\ell\cdot\textbf B^z$, so $\bar{\textbf B}^z=c\,\textbf B^z$ with $c=\sqrt3/2$ (kagome, $w_\ell=\sqrt3$) and $c=1/2$ (checkerboard, $w_\ell=1$).  At rank 2 the texture is demodulated at $+\bm K$ on each sublattice with the same kernel at $\sigma=1$,
\begin{equation}
\label{eq:supp-rank2-valley-projection}
 \bar\psi_a(\bR)=
 \frac{\sum_{i\in a}w_\sigma(\bR-\br_i)e^{-\ii\bm K\cdot(\br_i-\br_0)}\delta S^z_i}
 {\sum_{i\in a}w_\sigma(\bR-\br_i)},
 \qquad
 \bar\Phi_c=\bar\Phi_1+\ii\bar\Phi_2=\sqrt2\,\bar\psi_1,
\end{equation}
with $\br_0$ the centre of $\hexagon_0$ and the soft modes from Eq.~\eqref{eq:supp-rank2-field-reconstruction}: the reconstructed field is the demodulated amplitude of sublattice 1, and $\bar B_c=-\sqrt\lambda\,\bar\Phi_c$.  The conjugate valley, at $-2\bm K\equiv\bm K$, is suppressed by $e^{-\sigma^2|\bm K|^2/2}\simeq10^{-4}$, and the uniform background has no weight at $\bm K$.  The kernels only remove lattice-scale oscillations; they carry no phase and do not define the field.

On $N_\theta=160$ points $\bR_j=\br_0+R(\cos\theta_j,\sin\theta_j)$ of circles of integer radius $R\le L/3$ around the orphan, with $z_j=\bar B^z_x(\bR_j)+\ii\bar B^z_y(\bR_j)$ at rank 1 and $z_j=\bar\Phi_c(\bR_j)$ at rank 2, the defect winding of Eq.~\eqref{eq:defect-winding-prl} of the main text is evaluated as
\begin{equation}
\label{eq:supp-mc-estimator}
 W_{\rm MC}(R)=\frac{1}{2\pi}\sum_{j=0}^{N_\theta-1}\operatorname{Arg}\big[z_{j+1}z_j^*\big],
 \qquad z_{N_\theta}\equiv z_0 ,
\end{equation}
an integer whenever successive phases differ by less than $\pi$; a constant source or valley phase drops out.  The rank-1 envelope is the angular average of the radial flux $B_r=\bar{\textbf B}\cdot\hat{\br}$, and the tangential flux $B_\theta=\bar{\textbf B}\cdot\hat{\bm\theta}$ tests radiality.  The rank-2 envelope is the $m=2$ harmonic,
\begin{equation}
\label{eq:supp-rank2-harmonic}
 c_2(R)=\frac1{N_\theta}\sum_j\bar\psi_1(\bR_j)\,e^{-2\ii\theta_j},
 \qquad
 \mathcal F(R)=\frac{3\pi\,|c_2(R)|}{A_{\rm uc}\,q_{\rm core}\langle n^z\rangle},
 \qquad
 q_{\rm core}=m_{\rm O}-\frac{\langle C^z_0\rangle_{{\rm ud},h}}{\langle n^z\rangle}=\frac{Q_{\rm core}}{\langle n^z\rangle},
\end{equation}
normalized by Eq.~\eqref{eq:supp-rank2-derived-texture} with $Q_{\rm eff}\to Q_{\rm core}$, so that $\mathcal F=4\pi|B^z_c|_{m=2}/Q_{\rm core}=(T/\sigma_0^2)F_2(R/\xi_2)$ tends to $T/\sigma_0^2\le1.017$ in the Coulomb window; the phase of $c_2$ is the constant $\chi$.  The gauge charge is measured on the lattice without kernels, by summing the staggered vertex charge of Eq.~\eqref{eq:supp-general-incidence-charge} at rank 1, or the valley component of the constrainer of the texture, $\delta C^z_{\hexagon}=\sum_ic_{\hexagon i}\delta S^z_i$, at rank 2, over a disc around the orphan:
\begin{align}
\label{eq:supp-enclosed-charge}
 Q_{\rm enc}(R)&=\sum_{|\br_v-\br_0|\le R}\langle Q_v^z\rangle
 =Q_{\rm eff}\,\frac{R}{\xi_1}\,K_1\!\left(\frac{R}{\xi_1}\right),\\
\label{eq:supp-rank2-enclosed-charge}
 Q_{\bm K}(R)&=\operatorname{Re}\!\sum_{|\br_{\hexagon}-\br_0|\le R}e^{-\ii\bm K\cdot(\br_{\hexagon}-\br_0)}\,\delta C^z_{\hexagon}
 =Q_{\rm eff}\,G_2\!\left(\frac R{\xi_2}\right),
\end{align}

\begin{figure}[t!]
\centering
\includegraphics[width=\textwidth]{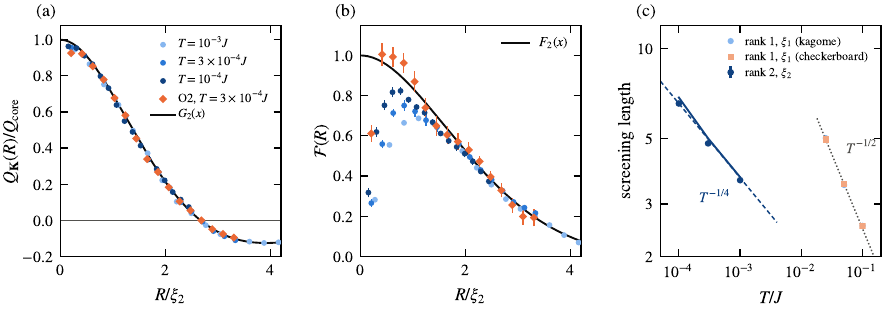}
\caption{Rank-2 screening ($L=48$, uniform field $h=0.2J$).  (a) Enclosed valley charge, Eq.~\eqref{eq:supp-rank2-enclosed-charge}, per unit charge $q_{\rm core}\langle n^z\rangle=Q_{\rm core}$, against $R/\xi_2$, for orphan O1 at three temperatures (light to dark) and orphan O2 at $T=3\times10^{-4}J$ (diamonds); the line is $G_2(x)$.  (b) Normalized envelope $\mathcal F$, Eq.~\eqref{eq:supp-rank2-harmonic}, against $R/\xi_2$ for the same data; the line is the continuum screening function $F_2(x)$ of Eq.~\eqref{eq:supp-rank2-texture}.  The points fall below it only in the core, $R\lesssim3$, where the kernel averages over a rapidly turning phase.  (c) Screening lengths from the charge fits: rank 2 (dark circles, SCGA line, dashed $T^{-1/4}$ guide) and, for comparison, rank 1 from Table~\ref{tab:mc-summary} (light symbols, dotted $T^{-1/2}$ guide).}
\label{fig:S-rank2-scaling}
\end{figure}
the continuum forms being the integrated screening clouds of Eqs.~\eqref{eq:supp-screening-cloud} and \eqref{eq:supp-rank2-G2}.  Because the constrainer vanishes at $\bm K$, a site all of whose snowflakes lie inside the disc does not contribute to $Q_{\bm K}$, so the convention $\delta S^z_i=0$ on vacancy sites drops out for $R\gtrsim3$.  Fits of these forms for $2\le R\le L/3$ give $Q_{\rm eff}$ and $\xi_n$ independently of the kernel width; a fit of the rank-1 envelope to $mK_1(mR)$ would absorb the smoothing.  For the errors, the blocks of each data set are grouped into 16 groups of at least $2.5\times10^4$ sweeps, much longer than the autocorrelation time, and every estimator is evaluated on group-averaged fields, with delete-one-group jackknife errors.  At rank 1 the reference blocks are grouped likewise, and each resample removes one defect group and one reference group together and recomputes $\langle n^z\rangle$ and $m_h$; at rank 2 the reference error, $10^{-7}$, and the relative error of $\langle n^z\rangle$, $2\times10^{-5}$, are negligible.  $Q_{\rm eff}$ and $\xi_n$ are refitted in every resample, so that their errors include the correlations between radii; at rank 1 this enlarges the errors by factors of up to three relative to a fit that treats the radii as independent.  A radius counts as robust when all 16 resamples return the integer of the full average (15 of 16 would be $94\%$); this measures the stability of the integer under resampling and is not a confidence level.  Radii at which the field reaches the noise floor fail this test and are shown as open symbols.

\begin{figure}[t!]
\centering
\includegraphics[width=0.85\textwidth]{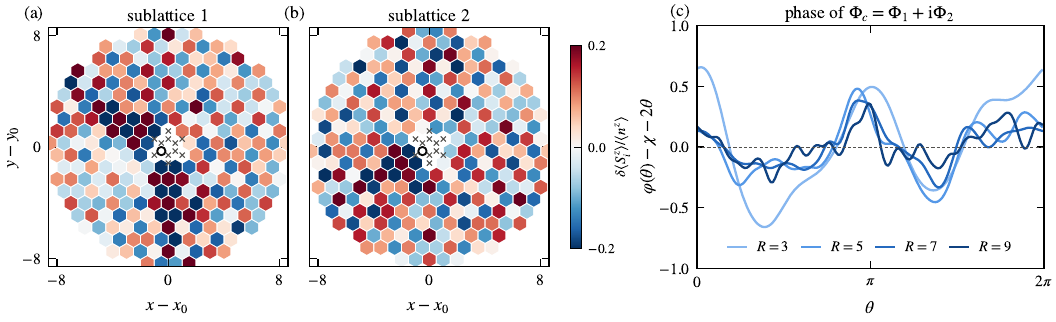}
\caption{Rank-2 texture ($L=48$, $T=10^{-4}J$, uniform field $h=0.2J$).  (a)~Measured magnetization relative to the reference run, per unit orphan magnetization, on sublattice 1, the sublattice from which $\Phi_c=\sqrt2\,\bar\psi_1$ is built: a $\sqrt3\times\sqrt3$ density wave at $\bm K$ whose phase slips around the orphan.  (b)~The same on sublattice 2, whose demodulated amplitude $\bar\psi_2=(\bar\Phi_1-\ii\bar\Phi_2)/\sqrt2$ is the opposite chiral projection of Eq.~\eqref{eq:supp-rank2-chiral-fields} and winds by $-2$.  The open circle marks the orphan and crosses the vacancy cluster.  (c)~Phase $\varphi(\theta)$ of the reconstructed field $\Phi_c=\Phi_1+\ii\Phi_2$, minus the constant $\chi$ and $2\theta$, on circles of radius $R=3$, $5$, $7$, $9$.  The deviation stays bounded, so the phase advances by exactly $4\pi$ around each circle.}
\label{fig:S-rank2-texture}
\end{figure}

\begin{figure}[t!]
\centering
\includegraphics[width=\textwidth]{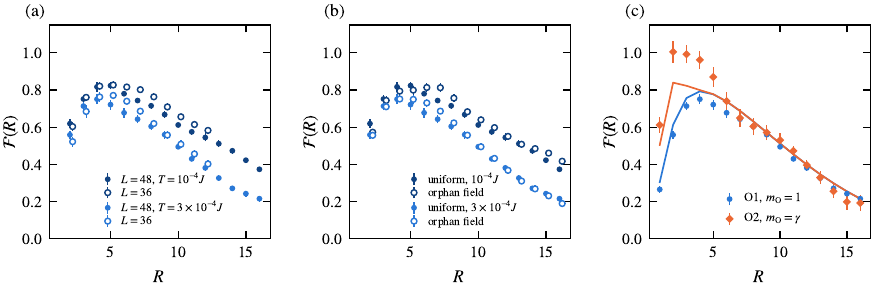}
\caption{Rank-2 checks (uniform field $h=0.2J$ unless stated).  (a) Normalized envelope at $L=48$ (filled) and $L=36$ (open) for $T=10^{-4}J$ (dark) and $3\times10^{-4}J$ (light).  (b) Envelope in the uniform field (filled) and in the orphan-field runs (open), each normalized by its own $q_{\rm core}$ ($L=48$; $T=10^{-4}J$ dark, $3\times10^{-4}J$ light).  (c) Envelope for the hexagon orphan O1 and the exterior orphan O2 ($L=48$, $T=3\times10^{-4}J$), each normalized by its own $q_{\rm core}$; lines are the SCGA.}
\label{fig:S-rank2-checks}
\end{figure}

The lines in Figs.~\ref{fig:rank1-mc-winding}(d), \ref{fig:rank2-mc-winding}(b) and \ref{fig:S-scaling}--\ref{fig:S-rank2-checks} are not fits.  The multiplier $\kappa$ follows from the spin-length condition on the undiluted lattice,
\begin{equation}
\label{eq:supp-kappa}
 \frac1{NN_{\rm uc}}\sum_{\bq}\sum_{b=1}^N\frac1{\kappa+\beta\,\omega_b(\bq)}=\frac13 ,
\end{equation}
with $\omega_b(\bq)$ the eigenvalues of $J_{ab}(\bq)$, zero on the flat bands (the signless Laplacian of the parent graph at rank 1; a $600\times600$ grid at rank 2); this gives $\kappa=1.04$--$1.60$ at rank 1 and $1.51$--$1.53$ at rank 2.  The texture is then the stationary point of the SCGA action $\frac12\textbf S^TM\textbf S-\beta h\sum_iS^z_i$, $M=\kappa\mathbb I+\beta J\,C^TC$ with $C$ the constrainer matrix of the lattice, subject to $S_i=0$ on every vacancy site and to $C_0=m_{\rm O}n^z$, each with its own multiplier $\mu_k$:
\begin{equation}
\label{eq:supp-lattice-saddle}
 \begin{pmatrix} M & E\\ E^T & 0\end{pmatrix}
 \begin{pmatrix}\langle\textbf S^z\rangle\\ -\bm\mu\end{pmatrix}
 =\begin{pmatrix}\beta h\,\textbf 1\\ \textbf g\end{pmatrix},
\end{equation}
where the columns of $E$ are the unit vectors of the vacancy sites and the coefficient vector of the orphan constrainer, and $\textbf g=(0,\dots,0,m_{\rm O}n^z)$.  The orphan is held at its measured magnetization, and the undiluted solution $M^{-1}\beta h\textbf 1$, the SCGA value of $m_h$, is subtracted.  This is Eq.~\eqref{eq:supp-full-multipliers} on the lattice.  Keeping only the last column of $E$ reproduces Eq.~\eqref{eq:texture-factorization-prl} of the main text, the last multiplier being the $\mu^z$ of Sec.~S2; the vacancy multipliers change the rank-1 texture by less than $3\%$ for $R\ge4$ (by $11$--$13\%$ at $R=2$) and the rank-2 envelope by less than $1\%$ for $R\ge3$, and leave the enclosed charge unchanged (Sec.~S2).  Passed through the same fits, the SCGA texture gives the charges $Q_{\rm eff}^{\rm SCGA}$ of Tables~\ref{tab:mc-summary} and \ref{tab:mc-rank2}; for $T\le0.05J$ they lie $1$--$3\%$ below the one-source $T\langle\mu^z\rangle$ ($6\%$ on kagome at $0.1J$), the systematic error of fitting the continuum forms to the lattice charge.  The sum of the texture per unit orphan magnetization, orphan included, is the SCGA value of the renormalized moment $\alpha$ of Sec.~S2.  The prediction is passed through the same estimators as the data.

\begin{table}[t!]
\centering
\small
\setlength{\tabcolsep}{6pt}
\begin{tabular}{llllllll}
\hline\hline
orphan & $T/J$ & $\alpha$ & $W_{\rm MC}=2$ for & $Q_{\rm eff}/\langle n^z\rangle$ & $q_{\rm core}$ & $Q_{\rm eff}^{\rm SCGA}/\langle n^z\rangle$ & $\xi_2$ \\
\hline
O1 & $10^{-4}$ & $0.226(2)$ & $R=2$--$16^{\,*}$ & $0.946(1)$ & 0.956 & 0.944 & $6.54(3)$ \\
O1 & $3\times10^{-4}$ & $0.224(3)$ & $R=2$--$15$ & $0.958(4)$ & 0.955 & 0.954 & $4.80(5)$ \\
O1 & $10^{-3}$ & $0.225(2)$ & $R=2$--$14$ & $0.951(5)$ & 0.955 & 0.952 & $3.61(2)$ \\
O2 & $3\times10^{-4}$ & $0.114(1)$ & $R=2$--$11$ & $0.455(3)$ & 0.455 & 0.455 & $4.84(7)$ \\
\hline\hline
\end{tabular}
\caption{Rank-2 data sets (honeycomb-snowflake model, $\gamma=1/2$, $h=0.2J$, $L=48$).  Columns as in Table~\ref{tab:mc-summary}, with $Q_{\rm eff}$ and $\xi_2$ from fits of Eq.~\eqref{eq:supp-rank2-enclosed-charge}; the charge per unit $\langle n^z\rangle$ is compared with $q_{\rm core}=Q_{\rm core}/\langle n^z\rangle$ of Eq.~\eqref{eq:supp-rank2-harmonic} and with the same fit applied to the lattice SCGA texture.}
\label{tab:mc-rank2}
\end{table}

\paragraph*{Rank-1 and rank-2 results.}
The data sets are summarized in Table~\ref{tab:mc-summary}.  The orphan follows $\langle n^z\rangle=L(\alpha h/T)$ with $\alpha$ within $0.004$ of the SCGA and close to the $T\to0$ value $1/2$ of Sec.~S2.  At $L=48$ and $T=0.05J$ the winding is $W_{\rm MC}=1$, robust in every jackknife resample, from $R=2$ to $R=11$ (kagome) and $12$ (checkerboard), more than three screening lengths, with a tangential flux consistent with zero [Fig.~\ref{fig:rank1-mc-winding}(c)]; the plateau is longer at $T=0.025J$ and persists at $L=32$.  At $T=0.1J$ it narrows to $R\approx3$--$7$: the vacancy cluster removes the magnetization $m_h$ of its links, a neutral but anisotropic $O(h)$ background that competes with the weak monopole in the core, and the orphan-field runs, which lack this background, recover $W_{\rm MC}=1$ out to $R=9$--$11$.  Figure~\ref{fig:S-scaling} shows the temperature dependence.  The radial envelope agrees with the SCGA within $8\%$ for $3\le R\le8$ at $T\le0.05J$ (a,b), the enclosed charge collapses onto $xK_1(x)$ at all temperatures (c), and $\xi_1\propto T^{-1/2}$ (d), with fitted exponents $0.51(2)$ (kagome) and $0.49(2)$ (checkerboard).  Table~\ref{tab:mc-summary} compares the fitted charge with $Q_{\rm core}$ and with $Q_{\rm eff}^{\rm SCGA}$, which contains the factor $T/\sigma_0^2$, with $\sigma_0^2/T=0.93$, $0.96$ and $0.98$ at $T/J=0.1$, $0.05$ and $0.025$ on both lattices, and all vacancy multipliers.  On checkerboard $Q_{\rm eff}$ lies within $3\%$ of $Q_{\rm eff}^{\rm SCGA}$, and within $1\%$ of $Q_{\rm core}$, for $T\le0.05J$ at $L=48$ ($4\%$ below $Q_{\rm eff}^{\rm SCGA}$ at $L=32$), and $13\%$ below it at $0.1J$, where the plateau is short.  On kagome it exceeds it by $14\%$ at $T=0.05J$, $8\%$ at $0.025J$ and $24\%$ at $0.1J$, an excess that decreases roughly in proportion to $T$.  Since the vacancy multipliers leave the SCGA charge unchanged (Sec.~S2), we attribute this excess to correlations beyond the Gaussian theory, which are strongest on kagome, whose clean structure factor also deviates most from the SCGA (Fig.~\ref{fig:S-structure-factor}); it affects neither the winding nor the scaling of $\xi_1$.  The envelope does not depend on the system size [Fig.~\ref{fig:S-checks}(a)], nor the winding on the coarse-graining width: for $\sigma=1.0$, $1.4$ and $1.8$ the robust plateau spans $R=2$--$9$, $2$--$11$ and $3$--$14$ on kagome and $2$--$10$, $2$--$12$ and $3$--$12$ on checkerboard.  Per unit $\langle n^z\rangle$ the uniform-field texture is proportional to the exact orphan-field texture of Eq.~\eqref{eq:supp-exact-factorization}, with slope $0.94$ set by the $O(h)$ charge shift [Fig.~\ref{fig:S-checks}(b)].

The data sets are summarized in Table~\ref{tab:mc-rank2}.  The renormalized moment is $\alpha\simeq0.225$ for O1 and $0.114$ for O2, in agreement with the SCGA and close to the $T\to0$ values $2/9$ and $1/9$: it is proportional to $m_{\rm O}$.  The raw magnetization is a $\sqrt3\times\sqrt3$ density wave at $\bm K$ whose phase slips around the orphan in opposite senses on the two sublattices [Fig.~\ref{fig:S-rank2-texture}(a,b)]: $\bar\psi_2$ winds by $-2$ on every circle $R=2$--$16$, as Eq.~\eqref{eq:supp-rank2-chiral-fields} requires.  The phase of $\Phi_c$ advances by $4\pi$ on every circle, with $\chi=\pi$, the sign of Eq.~\eqref{eq:supp-rank2-texture} [Fig.~\ref{fig:S-rank2-texture}(c)].  At $T=10^{-4}J$ the winding is $W_{\rm MC}=2$, robust in every jackknife resample, from $R=2$ to $R=16=L/3\approx2.4\,\xi_2$ [Fig.~\ref{fig:rank2-mc-winding}(c)], and the envelope lies within $8\%$ of the SCGA for $3\le R\le8$ [Fig.~\ref{fig:rank2-mc-winding}(b)].  Figure~\ref{fig:S-rank2-scaling} shows the screening: the enclosed valley charge collapses onto $G_2$, including its overscreened negative tail (a), the envelope collapses onto $F_2$ outside the core (b), and $\xi_2\propto T^{-0.26(1)}$ over the decade studied, consistent with the $T^{-1/4}$ of the continuum theory (the lattice SCGA gives an effective exponent of $0.27$ over the same range), against $T^{-1/2}$ at rank 1 (c).  The fitted charge agrees with $Q_{\rm eff}^{\rm SCGA}$ within $0.5\%$ for both orphans, including the noninteger $m_{\rm O}=\gamma$ of O2 and the $O(h)$ shift $\langle C^z_0\rangle_{{\rm ud},h}$, and with $Q_{\rm core}$ to about $1\%$, since $T/\sigma_0^2-1\le1.7\%$ at rank 2.  Figure~\ref{fig:S-rank2-checks} shows the checks.  The envelopes at $L=36$ and $48$ agree at $T=3\times10^{-4}J$; at $10^{-4}J$ the first four $L=36$ chains gave an envelope $10$--$15\%$ higher for $7\le R\le12$.  Twelve further independent starts, sixteen chains as at $L=48$, reduce the difference to $6$--$8\%$ (a), of which the SCGA accounts for $1$--$5\%$ as a finite-size effect of the torus; the chain-to-chain spread of the envelope is $7\%$ at $L=36$ and $5\%$ at $L=48$, and the remainder, about two standard errors, we attribute to the slow dynamics.  The winding and the charge are unaffected: $W_{\rm MC}=2$ for $R=2$--$12=L/3$ and $Q_{\rm eff}/\langle n^z\rangle=0.945(1)$ at $L=36$.  The winding does not depend on the kernel width: $W_{\rm MC}=2$ is robust from $R=2$ to $16$ for $\sigma=0.8$, $1.0$ and $1.4$.  The uniform-field and orphan-field envelopes agree within $9\%$ (b), and the envelopes of O1 and O2 differ only in the core, $R\lesssim5$ (c): the vacancy cluster sets the amplitude of the texture, and its shape only near the core.

\end{document}